\documentclass[english,eqsecnum,prd,aps,nofootinbib,superscriptaddress,longbibliography,tightenlines,11pt]{revtex4-2}
\usepackage[T1]{fontenc}
\usepackage[latin9]{inputenc}
\usepackage[dvipsnames]{xcolor}
\usepackage{babel}
\usepackage{mathtools}
\usepackage{amsmath}
\usepackage{amssymb,accents}
\usepackage{mathrsfs}
\usepackage{cancel}
\usepackage{hyperref}
\usepackage{verbatim}
\usepackage{subcaption}
\PassOptionsToPackage{normalem}{ulem}
\usepackage{ulem}

\begin{document}

\title{What do gravitational wave detectors say about polymer quantum effects?}

\author{Angel Garcia-Chung}
\email{alechung@xanum.uam.mx} 
\affiliation{Departamento de F\'isica, Universidad Aut\'onoma Metropolitana - Iztapalapa, \\ San Rafael Atlixco 186, Ciudad de M\'exico 09340, M\'exico.}

\affiliation{Tecnol\'ogico de Monterrey, Escuela de Ingenier\'ia y Ciencias, Carr. al Lago de Guadalupe Km. 3.5, Estado de Mexico 52926, Mexico.}

\author{Matthew F. Carney}
\email{c.matthew@wustl.edu}
\affiliation{Department of Physics and McDonnell Center for the Space Sciences,
Washington University, St. Louis, MO 63130, USA}

\author{James B. Mertens}
\email{jmertens@wustl.edu}
\affiliation{Department of Physics and McDonnell Center for the Space Sciences,
Washington University, St. Louis, MO 63130, USA}

\author{Aliasghar Parvizi}
\email{a.parvizi@ut.ac.ir}
\affiliation{Department of Physics, University of Tehran, North Karegar Ave., Tehran 14395-547, Iran.}
\affiliation{School of Physics, Institute for Research in Fundamental Sciences (IPM),
P.O. Box 19395-5531, Tehran, Iran}

\author{Saeed Rastgoo}
\email{srastgoo@ualberta.ca}
\affiliation{Department of Physics, University of Alberta, Edmonton, Alberta T6G 2E1, Canada}
\affiliation{Department of Mathematical and Statistical Sciences, University of Alberta, Edmonton, Alberta T6G 2G1, Canada}
\affiliation{Theoretical Physics Institute, University of Alberta, Edmonton, Alberta T6G 2E1, Canada}

\author{Yaser Tavakoli}
\email{yaser.tavakoli@guilan.ac.ir}
\affiliation{Department of Physics,
University of Guilan, Namjoo Blv.,
41335-1914 Rasht, Iran}
\affiliation{School of Astronomy, Institute for Research in Fundamental Sciences (IPM),	P. O. Box 19395-5531, Tehran, Iran}

\date{\today}
\begin{abstract}
We compute the expected response of detector arms of gravitational wave observatories to polymerized gravitational waves. The mathematical and theoretical features of these waves were discussed in our previous work. In the present manuscript, we find both perturbative analytical, and full nonperturbative numerical solutions to the equations of motion of the detector arms using the method of geodesic deviations. These results show the modifications to both frequency and amplitude of the signal measured by the detector. Furthermore, we study the detectability of these signals in LISA by analyzing the modes in the frequency space.
\end{abstract}

\maketitle


\section{Introduction}
We are living in the exciting era of multimessenger observatories where we are able to obtain signals from high energy phenomena via electromagnetic waves, neutrinos, and particularly, gravitational waves (GW). The most recent of these messengers, GWs, have opened up an unprecedented window of opportunity to study phenomena that could not be investigated experimentally prior to the discovery of these waves. Gravitational waves produced by the merger of compact objects [citation] have certainly revealed much about the properties of the objects that produce them, but these waves have the potential to reveal other aspects of the cosmos as well.
Perhaps the most exciting aspect of GWs for theoretical and fundamental physics is the possibility they provide for testing quantum gravity effects \citep{Addazi:2021xuf,LISA:2022kgy,LISACosmologyWorkingGroup:2022jok}. Indeed, the lack of experimental evidence for quantum gravity is and has been one of the most important issues in fundamental physics. However, recent and upcoming GW observatories such as LIGO, VIRGO, and LISA, give us the exciting possibility of finding phenomenological signatures of quantum gravity such as the quantum nature of spacetime, and potentially, the physics of the very early universe in the quantum gravity regime, to name a few. These instruments are thus very welcome additions to our multimessenger observatory arsenal and are crucial in advancement of the research in quantum gravity (for a nonexhaustive list of possibilities in phenomenology of quantum gravity with GWs see \cite{Amelino-Camelia:1998mjq,Giddings:2016tla,Arzano:2016twc,Addazi:2018uhd,Calcagni:2019ngc,Calcagni:2019kzo,Calcagni:2020tvw,Calcagni:2020ume,Bojowald:2007cd,Grain:2009eg,Dapor:2020jvc,Addazi:2018uhd,Barrau:2018rts,Maselli:2018fay,LISA:2022kgy,LISACosmologyWorkingGroup:2022jok}).

As mentioned above, GWs may be messengers by which we can study the possible fine/quantum structure of spacetime. This can be done either in a semiclassical regime or a fully quantum one. The semiclassical regime can be divided into two approaches. In the first semiclassical approach, the background spacetime over which the GW is propagating is quantized/discretized while the GW itself is considered as a classical wave. This classical wave, then, can be used to probe the fine structure of the (background) spacetime. In the second semiclassical approach, the background spacetime is classical while the GW is quantized. This is legitimate as a semiclassical approach since the GW is part of the metric or spacetime itself and hence its quantization may yield some information about the quantum nature of spacetime. This is the approach we use in this work and in our previous ones \citep{Garcia-Chung:2020zyq,Garcia-Chung:2021doi}. In fact, this method can also be used for the semiclassical approach in which Gamma Ray Bursts (GRB) propagate over spacetime \citep{Bonder:2017ckx}. A full quantum treatment can also be divided into two approaches. In the first approach both the background spacetime and the perturbations are quantized (while before quantization, one has divided the classical spacetime into a background and a perturbation). In the second approach, one first quantizes the whole spacetime nonperturbatively, obtains a semiclassical limit, and then in this limit divides the effective metric into a background and a foreground, and then studies the propagation of the latter on the former. This last approach is one we will consider in a future study.

There are many ways that one can quantize the background spacetime or the perturbations. In this work we use a quantization method, known as the polymer quantization. This is a nonperturbative method of quantization, in which usually either the configuration variable or its momentum does not admit a representation on a Hilbert space. Instead, a certain form of them resembling an exponential of the classical variables exists on this space. More precisely, on the Hilbert space, one of the canonical pairs is represented as the member of the algebra of the theory and the other one as the group member. This means that one loses the infinitesimal transformation in one of the variables and consequently the existence of only finite transformations leads to the discretization or quantization of the system (for more details and several examples, see e.g., \citep{Ashtekar:2002sn,Tecotl:2015cya,Morales-Tecotl:2016ijb}). 

Polymer quantization itself goes hand in hand with loop quantum gravity (LQG) \citep{Thiemann:2007pyv, Ashtekar:2004eh, Rovelli:2004tv} which is also a nonperturbative method of quantizing  the classical spacetime. There have been several studies exploring the potential observational prospects of the quantized spacetime structure in LQG \cite{Brizuela:2016gnz,Agullo:2015tca,Parvizi:2021ekr,Dapor:2020jvc,Garcia-Chung:2020zyq} 
Following what we described in previous paragraphs, our goal in this work is to predict LQG-inspired polymer effects that may be observed in GW detectors. In this approach, we consider a polymer quantized GW propagating on a classical spacetime and study the dynamics of the detector hands interacting with such a wave. 
More precisely, we assume that upon the creation of the GWs as perturbations in the gravitational field due to high energy phenomenon such as the merger of black holes, the quantum nature of spacetime leaves its imprints on the waves as LQG-polymer signatures. Mathematically this is translated into the Fourier modes of GWs being polymer quantized. Then, when these waves travel the large astronomical or cosmological distances towards our planet, their dynamics is governed by an effective polymer description. In both of the above stages the dynamics is governed by equations with non-linear corrections which depend on the polymer parameters. Once these GWs reach our detectors, interact with the detector's arms.  As mentioned above, in this work, we study the dynamics of the detector hands interacting with such waves. We will also analyze the detectability of these effects in LISA and will estimate the quantum gravity (or polymer) scale needed for such a detection.

This paper is organized as follows. In Sec.~\ref{sec:GRW-H} we present the Hamiltonian formalism of GWs in classical theory. We also give a brief introduction to polymer quantization and present the quantum Hamiltonian corresponding to the GWs and the associated effective Hamiltonian. We will review the effective theory for the wave propagation which we will need in the consequent sections. In Sec.~\ref{sec:phenomenology} we study the detector response to, and hence the resulting strain of, an effective polymerized GW. We first present the perturbative (in detector deviation) analytical solutions from which one can read off the modifications to the amplitude, frequency, and the speed of propagation of these effective GWs, up to the highest order in polymer (i.e., quantum gravity) scale. We then move on to the full nonperturbative (in detector deviation) solution to show that these effects are indeed nonperturbatively present, and no unexpected nonperturbative, secular type effects arise. 
In Sec.~\ref{sec:LISA} we analyze a black hole-black hole binary (BHB) system in our model and study the frequency space of the resulting signal. This allows us to  discuss the detectability of the polymer quantum gravity effects in such waves in LISA. Finally in Sec.~\ref{sec:Discussion} we conclude and discuss potential future directions.

\section{Hamiltonian formalism for GWs}
\label{sec:GRW-H}

In this section, we review the classical theory of GWs propagating on a flat spacetime background following \cite{Garcia-Chung:2020zyq}, polymer quantize this Hamiltonian, and compute the effective evolution equations for such waves.

\subsection{Classical theory}

GWs are the result of weak field approximation to Einstein's field
equations. On a flat background, these are generated by 
a small metric perturbation to the Minkowski background spacetime. Given the unperturbed Einstein-Hilbert gravitational action 
\begin{equation}
S_{\rm grav}\ =\ \frac{1}{2\kappa^2}\int d^4x \sqrt{-g}\, \mathcal{R} \, ,
\label{Eq:EH-Action}
\end{equation}
with $\kappa^2\equiv 8\pi G/c^4$, 
the general perturbed metric is written as
\begin{equation}
g_{\mu\nu} = \mathring{g}_{\mu\nu} +\, h_{\mu\nu}\, =\ \eta_{\mu\nu} +\, h_{\mu\nu}\, ,
\label{Eq:metric-pert}
\end{equation}
where $\mathring{g}_{\mu\nu}=\eta_{\mu\nu}$ is the background metric, in this case the Minkowski metric, 
and $h_{\mu\nu}$ denotes a small perturbation over $\eta_{\mu\nu}$. Moreover, we have
\begin{equation}
h^{\mu\nu}\, =\, \eta^{\mu\lambda}\eta^{\nu\tau}h_{\lambda\tau}.
\end{equation}
In order to reduce the number of terms in the linearized Einstein field equations, 
it is convenient to express the Einstein tensor 
in terms of the {\em trace-reversed} metric perturbation
\begin{equation}
\bar{h}_{\mu\nu}\, :=\, h_{\mu\nu} - \frac{1}{2}\eta_{\mu\nu}h\, ,
\end{equation}
where, 
$h=h^{~\mu}_{\mu}=\eta^{\mu\nu}h_{\mu\nu}$. 
Using the {\em Lorentz gauge}
\begin{equation}
\partial_{\mu}\bar{h}^{\mu\nu}=0,
\label{LorentzGauge}
\end{equation}
the linearized Einstein field equations in terms of  $\bar{h}_{\mu\nu}$ are expressed as a wave equation. Additionally,  by imposing the \emph{(synchronous) transverse-traceless} gauge 
\begin{equation}
\bar{h}=0,\quad \quad  \bar{h}_{0\mu}=0, \quad \quad {\rm and} \quad \quad \nabla_{i}\bar{h}^{ij}=0,
\end{equation}
the metric perturbation looks like a  transverse wave.
In other words, we consider only spatial, transverse and traceless perturbations propagating  on the unperturbed flat background. 

A wave traveling  along, say, the $x^{3}$ direction, can be separated into two polarizations of scalar modes $\bar{h}_{+}(x)$ and $\bar{h}_{\times}(x)$ as 
\begin{equation}
\bar{h}_{ij}(x) \, =\, \bar{h}_{+}(x) e_{ij}^{+} + \bar{h}_{\times}(x)e_{ij}^{\times},
\label{polarizedmetric1}
\end{equation}
where, 
\begin{align}
	e^{+}=  \left(\begin{array}{cc}
		1 & 0\\
		0 & -1
	\end{array}\right) \quad \quad \text{and}  \quad \quad 
	e^{\times}=  \left(\begin{array}{cc}
		0 & 1\\
		1 & 0
	\end{array}\right).
\end{align}
At second order in linear perturbation, in a traceless-transverse gauge, the perturbed action corresponding to this system becomes \citep{Bardeen:1980kt} 
\begin{align}
S_{\rm GW}\ \simeq  \frac{1}{8\kappa^2}\int d^4x  \sqrt{-\eta}\, 
\bar{h}_{ij}\, \mathring{\Box}\, \bar{h}^{ij} \, ,
\label{Eq:EH-Action2}
\end{align}
where $\mathring{\Box}\equiv \eta^{\mu\nu}\partial_\mu \partial_\nu$. The equations of motion associated to this action are,
\begin{equation}
\mathring{\Box}\, \bar{h}_{ij}(x) =0.
\label{eq:GWs}
\end{equation}
By substitution the Eq.~(\ref{polarizedmetric1}) into the perturbed action (\ref{Eq:EH-Action2}), 
the Lagrangian density at second order in linear perturbations
becomes
\begin{equation}
{\cal L}_{\check{h}}=\frac{1}{2}\sum_{\lambda =+,\times} \check{h}_{\lambda}\mathring{\Box} \check{h}_{\lambda}+{\cal O}(\check{h}_\lambda^{2}),
\label{eq:Lagrangian-Perturb}
\end{equation}
where,
\begin{equation}
    \check{h}_\lambda(x) \coloneqq \frac{\bar{h}_\lambda(x)}{2\kappa}\, .
\end{equation}
The effective action  of the independent polarization modes, provided by the Lagrangian density \eqref{eq:Lagrangian-Perturb}, is that of two massless scalar fields.
Thus, the equations of motion for the (scalar) perturbation $\check{h}_{\lambda}(x)$, with a fixed  $\lambda$, is simply the familiar Klein-Gordon equation,
\begin{equation}
\mathring{\Box}\, \check{h}_{\lambda}(x) =0.
\label{Eq:Field}
\end{equation}
Our aim henceforth, will be to study the polymer quantum theory of scalar perturbations $\check{h}_{\lambda}(x)$ --satisfying the Klein-Gordon equation (\ref{Eq:Field})-- propagating on a flat spacetime.

From the Lagrangian density \eqref{eq:Lagrangian-Perturb}, one can obtain the momentum $ \check{\pi}_{\lambda}$ conjugate to the  field $\check{h}_{\lambda}(x)$.
The classical solutions of the equation of
motion (\ref{Eq:Field}) can then be expanded on a spatial hypersurface $x^0=$constant, in Fourier modes as 
\begin{subequations}
	\label{eq:H-lambda0}
\begin{align}
\check{h}_{\lambda}(x^0,\mathbf{x})\,  &=\, \frac{1}{\ell^{3/2}}\sum_{\mathbf{k}\in\mathscr{L}}\mathfrak{h}_{\lambda,\mathbf{k}}(x^0)e^{i\mathbf{k}\cdot\mathbf{x}},\label{eq:H-lambda}\\
\check{\pi}_{\lambda}(x^0,\mathbf{x})\, &=\,  \frac{1}{\ell^{3/2}}\sum_{\mathbf{k}\in\mathscr{L}}\Pi_{\lambda,\mathbf{k}}(x^0)e^{i\mathbf{k}\cdot\mathbf{x}},\label{eq: pi-lambda}
\end{align} \label{eq:lambda-tot}
\end{subequations}
where the wave vector $\mathbf{k}=(k_{1},k_{2},k_{3})\in(2\pi\mathbb{Z}/\ell)^{3}$ spans to a
three-dimensional lattice $\mathscr{L}$ \citep{Ashtekar:2009mb}. We  assume that the  allowed Fourier components are those with the wavevectors in the reciprocal space of  an elementary cubical cell $\mathcal{V}$, equipped with coordinates $x^j\in(0, \ell)$, and denote by $V_o=\ell^3$ the volume of the $\mathcal{V}$.  Then, all integrations in the Fourier expansion will be restricted to this volume. This assumption helps us overcome the factitious infinities that will
arise in $\mathbb{R}^3$ topology in integrals due to infinite volumes. In other words, it naturally gives us a theoretical infrared cutoff in our framework, although it makes our results cutoff-dependent, but it is physically relevant for the gravitational system we are going to study in this paper, at the end, we will have freedom to choose the scale of cutoff regarding the given system under study. The advantage of choosing a three-dimensional lattice $\mathscr{L}$ is to avoid the discussion of boundary conditions for the fields. More precisely, it is an assumption on boundary conditions and compactness of the sources, which simplifies calculations. In another study, we will use our framework in the context of inflation and extend our analysis slightly to incorporate the $\mathbb{R}^3$ topology. 
  
The Fourier coefficients
are canonically conjugate satisfying the commutation relations $\{\mathfrak{h}_{\lambda,\mathbf{k}},\Pi_{\lambda,\mathbf{k}^{\prime}}\}=\delta_{\mathbf{k},-\mathbf{k}^{\prime}}$.
Moreover, the reality conditions on the field $\check{h}_{\lambda}(x^0,\mathbf{x})$  imply that
$\mathfrak{h}_{\lambda,\mathbf{k}}=\mathfrak{h}^{\ast}_{\lambda,-\mathbf{k}}$
and $\Pi_{\lambda,\mathbf{k}}=\Pi^{\ast}_{\lambda,-\mathbf{k}}$ 
are satisfied for each mode.
These conditions further indicate that not all the modes $\mathfrak{h}_{\lambda,\mathbf{k}}$
of the GWs are independent. In other words, when decomposing each
field mode $\mathfrak{h}_{\lambda,\mathbf{k}}$ and its conjugate
momentum $\Pi_{\lambda,\mathbf{k}}$ as,
\begin{subequations}
\begin{align}
\mathfrak{h}_{\lambda,\mathbf{k}} &\coloneqq  \frac{1}{\sqrt{2}}\big(\mathfrak{h}_{\lambda,\mathbf{k}}^{(1)}+i\mathfrak{h}_{\lambda,\mathbf{k}}^{(2)}\big),\label{app-phi-pi-1a}\\
\Pi_{\lambda,\mathbf{k}} &\coloneqq  \frac{1}{\sqrt{2}}\big(\Pi_{\lambda,\mathbf{k}}^{(1)}+i\Pi_{\lambda,\mathbf{k}}^{(2)}\big),\label{app-phi-pi-1b}
\end{align}
\end{subequations}
the reality conditions on $\mathfrak{h}_{\lambda,\mathbf{k}}$ and $\Pi_{\lambda,\mathbf{k}}$ enable
us to split the lattice $\mathscr{L}$ into positive and negative sectors $\mathscr{L}_{+}$ and $\mathscr{L}_{-}$, respectively.
Thereby, any summation over $\mathbf{k}\in\mathscr{L}$ can be decomposed into
its positive (for $\mathbf{k}\in \mathscr{L}_{+}$) and negative (for $\mathbf{k}\in \mathscr{L}_{-}$) parts. 
Associated with these separate sectors, we can now define new variables ${\cal A}_{\lambda,\mathbf{k}}$
and ${\cal E}_{\mathbf{\lambda,k}}$ as 
\begin{subequations}
\begin{align}
{\cal A}_{\lambda,\mathbf{k}} &\coloneqq \begin{cases}
\mathfrak{h}_{\lambda,\mathbf{k}}^{(1)} & \textrm{for}\quad\mathbf{k}\in\mathscr{L}_{+};\\
\mathfrak{h}_{\lambda,-\mathbf{k}}^{(2)} & \textrm{for}\quad\mathbf{k}\in\mathscr{L}_{-},
\end{cases}\label{def-q}\\
{\cal E}_{\mathbf{\lambda,k}} &\coloneqq  \begin{cases}
\Pi_{\lambda,\mathbf{k}}^{(1)} & \textrm{for}\quad\mathbf{k}\in\mathscr{L}_{+};\\
\Pi_{\lambda,-\mathbf{k}}^{(2)} & \textrm{for}\quad\mathbf{k}\in\mathscr{L}_{-},
\end{cases}\label{def-p}
\end{align}
\end{subequations}
which are canonically conjugate,
\begin{equation}
\left\{ {\cal A}_{\lambda,\mathbf{k}},{\cal E}_{\mathbf{\lambda^{\prime},k}^{\prime}}\right\} =\delta_{\mathbf{k}\mathbf{k}^{\prime}}\delta_{\lambda\lambda^{\prime}}.
\label{eq:PB-AE}
\end{equation}
Using the Lagrangian \eqref{eq:Lagrangian-Perturb},  we can now express the Hamiltonian of the perturbation field, in terms of the new variables (\ref{def-q}) and (\ref{def-p}),  as 
\begin{align}
H\, &=\, \frac{1}{2}\sum_{\lambda=+,\times}\sum_{\mathbf{k}\in\mathscr{L}}\left[{\cal E}_{\mathbf{\lambda,k}}^{2}+k^{2}{\cal A}_{\lambda,\mathbf{k}}^{2}\right] \nonumber \\
\, &\eqqcolon\, \sum_{\lambda=+,\times}\sum_{\mathbf{k}\in\mathscr{L}} H_{\lambda, \mathbf{k}},
\label{eq:Hamiltonian-FLRW-2}
\end{align}
where $k=|\mathbf{k}|$.
Eq.~\eqref{eq:Hamiltonian-FLRW-2}  represents the Hamiltonian
of a set of decoupled harmonic oscillators defined by conjugate pairs
$({\cal A}_{\lambda,\mathbf{k}},{\cal E}_{\mathbf{\lambda,k}})$ associated
with a mode $\mathbf{k}$ for a fixed polarization $\lambda$, and satisfying the relation \eqref{eq:PB-AE}. In the next subsection we will provide the effective polymer Hamiltonian associated with the above classical Hamiltonian.

\subsection{Polymer quantum theory: Effective dynamics\label{PolySubSection}} 

The polymer quantization of the Hamiltonian \eqref{eq:Hamiltonian-FLRW-2} requires three main ingredients: (i) the Weyl algebra of quantum observables, (ii) the polymer Hilbert space together with the representation of the observables, and (iii) the polymer analog of the momentum operator. The first two ingredients are rather natural for many quantum descriptions (more details about the Weyl algebra is provided further below) but the third ingredient requires some clarification. 

Polymer quantum mechanics and loop quantum cosmology (LQC) are very similar quantization schemes at the mathematical level. They are singular representations of the Weyl algebra, in the sense that the quantum states cannot be transformed under infinitesimal translations due to the fact that the associated generators do not exist on the Hilbert space. The Hilbert space only admits the finite generators of such transformations. Because the momentum operator in a mechanical system is usually the generator of infinitesimal translations, these quantization schemes do not provide a representation for the momentum operator. That is why the third ingredient is needed. More details of the polymer quantization and its relation with LQC and LQG can be found in the literature \cite{Ashtekar:2002sn, CorichiVZ, VelhinhoJM, GarciaChung}.

As we mentioned before, the first ingredient is the Weyl algebra of quantum observables. In this context, the Weyl algebra is the set of abstract operators whose multiplication contains the canonical commutation relations but in exponential form, sometimes denoted as
\begin{align} \label{WeylAlgebraMultiplication}
\widehat{W}(a_1,b_1) \widehat{W}(a_2,b_2)= e^{- \frac{i}{2\hbar}(a_1 b_2- b_1 a_2)} \widehat{W}(a_1+a_2, b_1+b_2).
\end{align}
Here, the elements $\widehat{W}(a,b)$ denotes each of the generator of the Weyl algebra labelled with $a,b \in \mathbb{R}$ and are formaly given as
\begin{align}\label{WeylAlgebraGenerator}
\widehat{W}(a,b) := \widehat{e^{\frac{i}{\hbar}(a x + b p)}}.
\end{align}
Note that the operator symbol (hat) is acting over the entire exponential instead of the functions $x$ or $p$. With this, we imply that it is the entire exponential function what should be considered as an abstract operator. Historically, the linear form of the canonical commutation relations, sometimes denoted as e.g.,  $[\widehat{x}, \widehat{p}] = i \hbar$, is more familiar. However, it is not suitable to explore a possible discrete nature of the space because a discrete space forbids the standard notions of infinitesimal translations. This is the main reason to consider the Weyl algebra in polymer quantum mechanics.

On the other hand, the approach we follow is one in which only one of the fundamental operators will have discrete eigenvalues, whether the position operator or the momentum operator. As a result, in Eq.(\ref{WeylAlgebraGenerator}) instead of considering the entire Weyl algebra generator with the labels $a, b \neq 0$ we can take $a=0$ or $b=0$ and denote the resulting generator as
\begin{align}
\widehat{W}(a,0) = \widehat{V}(a), \qquad \widehat{W}(0,b) = \widehat{U}(b),
\end{align}
\noindent and the canonical commutation relations in Eq.~(\ref{WeylAlgebraMultiplication}) take the form 
\begin{align}
\left[ \widehat{U}(b), \widehat{x} \right] = \hbar \, b \, \widehat{U}(b), \qquad \mbox{or} \qquad \left[ \widehat{p}, \widehat{V}(a) \right] = \hbar \, a \, \widehat{V}(a). 
\end{align}
Now that we clarified the main aspects regarding the Weyl algebra structure we are ready to adapt these mathematical description to our current model for the canonical variables describing the gravitational waves.

We will consider two cases in this work and refer to them as polarizations. The first case, which we call ``polymer ${\cal E}_{\mathbf{\lambda,k}}$'', is where there is no infinitesimal operator ${\cal E}_{\mathbf{\lambda,k}}$ and the operator ${\cal A}_{\mathbf{\lambda,k}}$ has discrete eigenvalues. The second polarization, called ``polymer ${\cal A}_{\lambda,\mathbf{k}}$'', is the case where no infinitesimal operator ${\cal A}_{\lambda,\mathbf{k}}$ exists and the eigenvalues of ${\cal E}_{\mathbf{\lambda,k}}$ are discrete.

The observables in the polymer ${\cal E}_{\mathbf{\lambda,k}}$ case are given by $\widehat{{\cal A}}_{\lambda,\mathbf{k}}$ and $\widehat{U}_{\mathbf{\lambda,k}}(\mu)$. Here, the operator $\widehat{U}_{\mathbf{\lambda,k}}(\mu)$ is the generator of finite (discrete) translations. Note that when considered in the standard Schr\"odinger representation, this operator resembles the exponential of the momentum operator. The parameter $\mu$ is a dimensionful parameter encoding the discreteness of the operator $\widehat{{\cal A}}_{\lambda,\mathbf{k}}$. In this context, since $\widehat{{\cal A}}_{\lambda,\mathbf{k}}$ is related to the perturbation of the metric tensor, the parameter $\mu$ is thus associated with the discreteness of the spacetime. The commutation relation for these operators reads
\begin{align}
\left[ \widehat{U}_{\mathbf{\lambda,k}}(\mu), \widehat{{\cal A}}_{\lambda,\mathbf{k}} \right] =  \hbar \, \mu \, \widehat{U}_{\mathbf{\lambda,k}}(\mu).
\end{align}

The observables for the polymer ${\cal A}_{\lambda,\mathbf{k}}$ case are given by $\widehat{{\cal E}}_{\mathbf{\lambda,k}}$ and $\widehat{V}_{\lambda,\mathbf{k}}(\nu) $. As mentioned, in this case the eigenvalues of $\widehat{{\cal E}}_{\mathbf{\lambda,k}}$ are discrete. Analogous to the previous case, the parameter $\nu$ is the polymer scale related to the discreteness of the canonical conjugate momentum to the metric. The commutator for these observables is of the form
\begin{align}
\left[ \widehat{V}_{\lambda,\mathbf{k}}(\nu), \widehat{{\cal E}}_{\mathbf{\lambda,k}} \right] =- \hbar \, \nu \, \widehat{V}_{\lambda,\mathbf{k}}(\nu).
\end{align}

Although the representation of these operators is given in two different Hilbert spaces, they are very similar at the mathematical level. The Hilbert spaces for the polymer ${\cal E}_{\mathbf{\lambda,k}}$ and the polymer ${\cal A}_{\mathbf{\lambda,k}}$ polarizations are given, respectively, by
\begin{align}
{\cal H}_{{\rm poly}\, {\cal E}} = L^2(\mathbb{R}_d, d {\cal A}_c ), \qquad {\cal H}_{{\rm poly}\, {\cal A}} = L^2(\mathbb{R}_d, d {\cal E}_c ),
\end{align}
In both cases the configuration spaces are given by the real line with discrete topology, denoted by $\mathbb{R}_d$, and the measure is given by the countable measure on these discrete real lines. This results in a violation of the Stone-von Neumann theorem conditions \cite{Ashtekar:2002sn, VelhinhoJM, CorichiVZ}, and yields a polymer Hilbert space unitarily inequivalent to the usual Hilbert space of standard Schr\"odinger representation. Consequently, the polymer theory gives rise to new physics compared to the standard quantization scheme. Since the standard quantum mechanics predictions fit very well with the experiments of the systems with finite degrees of freedom, the predictions of the polymer quantum mechanics should also fit very well with the experiments for those systems. This criterion leads to bounds on the scale at which the polymer effects take over which in practice means bounds on the polymer scales parameters $\mu$ or $\nu$. Clearly, such a restriction is
not required in the polymer quantization of GWs where 
no measurement on the
quantum nature of GWs have been performed.

As usual, there is always a margin
of error or discrepancy between the theory and experiment, even if
the theory and experimental results match to a very high degree of
accuracy. This leaves room for possible new physics. In our case this
new physics is the polymer quantum theory. Hence it is worth investigating
the polymer quantum mechanics predictions for the GWs detectors, in
the hope that for certain values of the polymer parameters, the predictions
of the polymer model fit with the data to a degree higher than that
provided by the standard non-polymer models.

At the core of our analysis lays the assumption that the polymer effects
are small deviations when compared to the main contributions described
by the standard non-polymer models. Such models fit, to a high degree
of accuracy (more than $5\sigma$), with the detectors data using
the classical description of the Fourier modes of the GWs. Hence,
we expect $\mu$ or $\nu$ to be small such that in the limit, when
$\mu,\nu\rightarrow0$, we recover the contributions of the standard
quantum mechanical models. Since the polymer parameters have to be
considered very small, an effective description provides the minimum
insight we need to begin with. In other words, instead of moving towards
the full quantum polymer description we move directly to the effective
(classical) description already presented in \cite{Garcia-Chung:2020zyq}.

In this effective description, the Hamiltonian of each mode and polarization
in Eq. \eqref{eq:Hamiltonian-FLRW-2} is modified in order to incorporate
the first order contribution of the polymer parameters. This results
in two polymer effective (classical) Hamiltonians, one for each of
the two representations of the polymer model. The polymer ${\cal E}_{\mathbf{\lambda,k}}$
Hamiltonian is of the form 
\begin{align}
H_{\lambda,{\bf k}}^{({\cal E})}=\frac{2}{\mu^{2}}\sin^{2}\left(\frac{\mu\,{\cal E}_{\mathbf{\lambda,k}}}{2}\right)+\frac{1}{2}{\bf k}^{2}\,{\cal A}_{\mathbf{\lambda,k}}^{2},\label{PolymerE}
\end{align}
whereas the Hamiltonian for the polymer ${\cal A}_{\mathbf{\lambda,k}}$
is 
\begin{align}
H_{\lambda,{\bf k}}^{({\cal A})}=\frac{1}{2}{\cal E}_{\mathbf{\lambda,k}}^{2}+\frac{2}{\nu^{2}}\sin^{2}\left(\frac{\nu\,{\cal A}_{\mathbf{\lambda,k}}}{2}\right).\label{PolymerA}
\end{align}
Using these Hamiltonians, we can find the equations of motion (EoM)
as usual using Poisson brackets in each case (i.e., the polymer $\mathcal{E}$
and polymer $\mathcal{A}$ cases). We summarize these equations and
their solutions (without loss of generality only for the $+$ polarization)
in the following:   
\begin{itemize}
    \item[a)]  In polymer $\mathcal{E}$ case, the EoM read 
\begin{align}
\frac{d{\cal A}_{+,\mathbf{k}}}{dt}= & \frac{\hbar}{\mu}\,\sin\left(\frac{\mu}{\hbar}\,\mathcal{E}_{+,\mathbf{k}}\right), & \frac{d{\cal E}_{+,\mathbf{k}}}{dt}= & -k^{2}\mathcal{A}_{+,\mathbf{k}}.\label{eq:EoM-eff-E-2}
\end{align}
Combining these, the second order effective EoM for the GWs become 
\begin{equation}
\ddot{\mathcal{A}}_{+,\mathbf{k}}=-k^{2}\mathcal{A}_{+,\mathbf{k}}\cos\left(\frac{\mu}{\hbar}\,\mathcal{E}_{+,\mathbf{k}}\right).\label{eq:EoM-eff-E-nonH}
\end{equation}
Now we consider a situation in which the $(\mu/\hbar)\mathcal{E}_{+,\mathbf{k}}$
is small. By expanding the sine and cosine functions up to order $\mathcal{O}((\mu\mathcal{E}_{+,\mathbf{k}})^{2}/\hbar)$,
and after the re-scaling $\mathcal{A}\to(\hbar/\mu)\bar{\mathcal{A}}$,
Eq.~(\ref{eq:EoM-eff-E-nonH}) is approximated by 
\begin{equation}
\ddot{\bar{\mathcal{A}}}_{+,\mathbf{k}}+k^{2}\bar{\mathcal{A}}_{+,\mathbf{k}}\,\approx\,\dfrac{k^{2}}{2}\bar{\mathcal{A}}_{+,\mathbf{k}}\,\dot{\bar{\mathcal{A}}}_{+,\mathbf{k}}^{2}\,.\label{eq:EoM-eff-E-pert}
\end{equation}
By using the Poincare-Lindstedt method \cite{amore2005improved},
we can compute a perturbative solution without any secular (growing)
term at a given order. This yields
\begin{align}
\bar{\mathcal{A}}_{+,\mathbf{k}}^{(\mathcal{E})}(t)\approx & \bar{\mathcal{A}}_{I}\left[\left(1-\frac{\bar{\mathcal{A}}_{I}^{2}k^{2}}{32}\right)\cos\left(3kc\sqrt{1-\frac{\bar{\mathcal{A}}_{I}^{2}k^{2}}{8}}\,t\right)\right.\nonumber \\
 & \quad\quad\quad\quad\left.-\frac{\bar{\mathcal{A}}_{I}^{2}k^{2}}{64}\cos\left(3kc\sqrt{1-\frac{\bar{\mathcal{A}}_{I}^{2}k^{2}}{8}}\,t\right)\right].\label{eq:EoM-eff-E-sol-app1}
\end{align}
In terms of the original perturbation scalar, $\bar{h}(t)$, the above
solution is rewritten as 
\begin{align}
\bar{h}_{+,\mathbf{k}}^{(\mathcal{E})}(t)\approx & \bar{h}_{I}\left[\left(1-\frac{\bar{h}_{I}^{2}\bar{\mu}^{2}k^{2}}{32\hbar^{2}}\right)\cos\left(kc\sqrt{1-\frac{\bar{h}_{I}^{2}\bar{\mu}^{2}k^{2}}{8\hbar^{2}}}\,t\right)\right.\nonumber \\
 & \quad\quad\quad\quad\left.-\frac{\bar{h}_{I}^{2}\bar{\mu}^{2}k^{2}}{64\hbar^{2}}\cos\left(3kc\sqrt{1-\frac{\bar{h}_{I}^{2}\bar{\mu}^{2}k^{2}}{8\hbar^{2}}}\,t\right)\right],\label{eq:EoM-eff-E-sol-app2}
\end{align}
where we have defined a new polymer parameter $\bar{\mu}\equiv\mu\ell^{3/2}/2\kappa$
with the dimension of length. 
     \item[b)] For the polymer $\mathcal{A}$ case, the EoM derived from the Hamiltonian
\eqref{PolymerA} read
\begin{align}
\frac{d{\cal A}_{+,\mathbf{k}}}{dt}= & \mathcal{E}_{+,\mathbf{k}}, & \frac{d{\cal E}_{+,\mathbf{k}}}{dt}= & -\frac{\hbar k^{2}}{\nu}\,\sin\left(\frac{\nu}{\hbar}\mathcal{A}_{+,\mathbf{k}}\right),\label{eq:EoM-eff-A-2}
\end{align}
thereby, the second order effective EoM for $\mathcal{A}_{+,\mathbf{k}}$
becomes 
\begin{equation}
\ddot{\mathcal{A}}_{+,\mathbf{k}}+\frac{\hbar k^{2}}{\nu}\,\sin\left(\frac{\nu}{\hbar}\mathcal{A}_{+,\mathbf{k}}\right)=0.\label{eq:EoM-eff-A-nonH}
\end{equation}
For small $(\nu/\hbar)\mathcal{A}_{+,\mathbf{k}}$, equation above
up to $\mathcal{O}((\nu\mathcal{A}_{+,\mathbf{k}})^{3}/\hbar)$ becomes
\begin{equation}
\ddot{\bar{\mathcal{A}}}_{+,\mathbf{k}}+k^{2}\bar{\mathcal{A}}_{+,\mathbf{k}}=\frac{k^{2}}{6}\bar{\mathcal{A}}_{+,\mathbf{k}}^{3},\label{eq:EoM-eff-A-pert}
\end{equation}
where, again we have used the re-scaling $\mathcal{A}\to(\hbar/\nu)\bar{\mathcal{A}}$.

\noindent Using the Poincare-Lindstedt method, the solution to Eq.~(\ref{eq:EoM-eff-A-pert})
is approximated as 
\begin{equation}
\bar{\mathcal{A}}_{+,\mathbf{k}}(t)\,\approx\,\bar{A}_{I}\left(1-\frac{\bar{A}_{I}^{2}}{96}\right)\cos\left(kc\sqrt{1-\frac{\bar{A}_{I}^{2}}{8}}\,t\right)-\frac{\bar{A}_{I}^{3}}{192}\cos\left(3kc\sqrt{1-\frac{\bar{A}_{I}^{2}}{8}}\,t\right),\label{eq:EoM-eff-A-sol-app}
\end{equation}
which again, written in terms of the original variable $\bar{h}_{+,\mathbf{k}}$,
reads 
\begin{align}
\bar{h}_{+,\mathbf{k}}^{(\mathcal{A})}(t)\, & \approx\,\bar{h}_{I}\left[\left(1-\frac{\bar{h}_{I}^{2}\bar{\nu}^{2}}{96\,\hbar^{2}}\right)\cos\left(kc\sqrt{1-\frac{\bar{h}_{I}^{2}\bar{\nu}^{2}}{8\hbar^{2}}}\,t\right)\right.\nonumber \\
 & \quad\quad\quad\quad\left.-\frac{\bar{h}_{I}^{2}\bar{\nu}^{2}}{192\hbar^{2}}\cos\left(3kc\sqrt{1-\frac{\bar{h}_{I}^{2}\bar{\nu}^{2}}{8\hbar^{2}}}\,t\right)\right].\label{eq:EoM-eff-A-sol-app2}
\end{align}
Here, similar to the previous case we have defined a new dimensionless
(in natural units) polymer parameter $\bar{\nu}\equiv\nu\ell^{3/2}/2\kappa$. 
\end{itemize}

It is instructive to calculate the speed of GWs in both  polymer $\cal{A}$ and polymer $\cal{E}$ cases. To do so, we first calculate the dispersion relation of GWs including the polymer corrections. To the leading order these are given by
\begin{align}
\omega^{(\mathcal{A})} & \approx kc\left(1 - \frac{\bar{h}_I^2 \bar{\nu}^2}{8\hbar^2} \right)^{1/2}, \label{rel:disperionA}\\
\omega^{(\mathcal{E})} & \approx kc\left(1 - \frac{\bar{h}_I^2 \bar{\mu}^2 k^2}{8\hbar^2} \right)^{1/2}. \label{rel:disperionE}
\end{align}
The speed of propagation of GWs is given by the relation $v = d\omega/dk$. Thus, for our dispersion relations (\ref{rel:disperionA}) and (\ref{rel:disperionE}) we get
\begin{align}
v^{(\mathcal{A})} &\approx c \left(1 - \frac{\bar{h}_I^2 \bar{\nu}^2}{16\hbar^2} \right),  \\
v^{(\mathcal{E})} &\approx c \left(1 - \frac{3 \bar{h}_I^2 \bar{\mu}^2 }{16\hbar^2}\, k^2 \right). \label{rel:speedE}
\end{align}
The relation \eqref{rel:speedE} represents (in polymer $\cal{E}$ case) a phenomenological aspect of the effect of polymer quantization on the propagation of GWs. It shows that different modes of GWs will travel with different speeds under such effects. In particular, all the modes propagate subluminally, and the higher the energy of a mode, the lower its speed.

In the next section we will study another phenomenological aspect of the polymer GWs, namely, the effective geodesic deviation equation describing the detector's arms motion. This equation is coupled to the background perturbation whose dynamic is given by solutions  (\ref{eq:EoM-eff-E-sol-app2}) and (\ref{eq:EoM-eff-A-sol-app2}) and reveals the behavior of the detector's arms in this model.

\section{Effective dynamics of the arm length
	\label{sec:phenomenology}}

In this section, to investigate the consequence of polymer quantization on propagation of GWs, we will first present the geodesic deviation equation and find the effective evolution of the detector's arms. Using that, we will study the corresponding solutions to the detector's arm length, using analytical and numerical techniques.

\subsection{Geodesic deviation equation}
\label{sec:GeodesicDeviation}

The GW detector arms can be modeled as two free-falling (identical) masses whose geodesic separation  is sensitive to the Riemann tensor induced by gauge-invariant metric perturbations (or strain) $\mathcal{A}_{\lambda, \mathbf{k}}$, i.e., the incident GWs herein our setting. Geodesic equations of such  masses are given by the action \cite{maggiore2008gravitational}
\begin{align}
S_{\xi} 
&=  - m \int_{\gamma_{A}(t)}d\tau - m \int_{\gamma_{B}(t^{\prime})}d\tau^{\prime},  \nonumber\\
&= -m \int_{\gamma_{A}(t)} \sqrt{-g_{\mu\nu}\,dx^\mu\,dx^\nu} -m \int_{\gamma_{B}(t^{\prime})} \sqrt{-g^{\prime}_{\mu\nu}\,{dx^{\prime}}^\mu\,{dx^{\prime}}^\nu}\,, \label{def:action-Masses-1}
\end{align}
where, $\gamma_A(t)$ and $\gamma_B(t^{\prime})$ are timelike geodesics of the particles $A$ and $B$, respectively. We introduce a Fermi normal coordinates along the geodesics $\gamma_A(t)$ of particle $A$, 
situated at time $t$, at the point $P=(t,\mathbf{x}=0)$,
and whose geodesic is parameterized by time $t$. In such a frame,
the Fermi normal coordinates of particle $B$ (moving on geodesic
$\gamma_{B}$) are given by $\xi^{\mu}=(t,\xi^{i}(t))$ in the vicinity
of the point $P$. Thus $\xi^{i}$ represents the deviation parametrized
by $t$, i.e., $\xi^{i}$ connects two points with the same value
of $t$ on the two geodesics. In this configuration we can recast
the action \eqref{def:action-Masses-1} in terms of the deviation
variables and only focus on the particle with geodesic $\gamma_{B}$
while ignoring the dynamics of the particle with geodesic $\gamma_{A}$.
This is because, as is well-known, in these coordinates on the worldline
of particle $A$ we have $g_{\mu\nu}|_{\gamma_{A}}=\eta_{\mu\nu}$.
We then write the above action as
\begin{eqnarray}
S_{\xi}=-m\int_{\gamma_{B}}dt\sqrt{-g_{\mu\nu}^{\prime}\left(\xi^{i},t\right)\,\dot{\xi}^{\mu}\,\dot{\xi}^{\nu}}\,.\label{def:action-Masses-2}
\end{eqnarray}
around point $P$, the metric $g_{\mu\nu}^{\prime}$ in the neighborhood
of $\gamma_{A}$ can be expanded as \cite{2007reto.book.....P} 
\begin{align}
ds^{2}\simeq & -\left(1+R_{0i0j}\,\xi^{i}\xi^{j}+\mathcal{O}(\xi^{3})\right)dt^{2}-\left(\frac{4}{3}R_{0jik}\xi^{j}\xi^{k}+\mathcal{O}(\xi^{3})\right)\,dtdx^{i}\nonumber \\
 & +\left(\delta_{ij}-\frac{1}{3}R_{ikj\ell}\,\xi^{k}\xi^{\ell}+\mathcal{O}(\xi^{3})\right)dx^{i}dx^{j}.\label{def:metric-fermi}
\end{align}
In the proper detector frame on the Earth, the metric \eqref{def:metric-fermi} has other contributions coming from Newtonian forces such as, Newtonian gravity, Coriolis and centrifugal forces, suspension mechanism and Sagnac effect. These effects are many order of magnitude larger than GWs but change very slowly. To detect the GWs, we need to use a higher frequency window in which noises from other sources are very small and the main contribution in the metric \eqref{def:metric-fermi} comes from GWs, thus we can obtain the geodesic deviation induced mainly by GWs.
Replacing the metric \eqref{def:metric-fermi} into Eq.~\eqref{def:action-Masses-2}, the action for the geodesic deviation becomes,
\begin{equation}
S_{\xi} \simeq \int_{\gamma_{B}} dt \left[ \frac{m}{2} \dot{\xi}^{i^2} +  \frac{m}{4} \ddot{\bar{h}}_{ij} (\mathbf{0},t) \xi^i \xi^j \right] \, , \label{GDAction-Ali}
\end{equation}
where we have dropped the non-dynamical terms. To the first order in metric perturbations, in the TT gauge, we have $R_{0i0j}(t,\mathbf{0})=- \ddot{h}_{ij} (t,\mathbf{0})/2$, in which the Riemann tensor is evaluated at the point $P$. 

The Hamiltonian of the geodesic deviation can be obtained with the help of the Legendre transformation of the action \eqref{GDAction-Ali} as
\begin{align}
H(t) = \frac{1}{2m} \sum_{i=1}^{2}P^2_{\xi^i} - \frac{m}{4} \ddot{\bar{h}}_{ij} (\mathbf{0},t) \xi^i \xi^j, \label{eq:Hamiltonian-interaction}
\end{align}
leading to the EoM of the form
\begin{align}
\dot{\xi}_i &= \frac{1}{m} P_{\xi^i},\label{HamiltonEqs0} \\
\dot{P}_{\xi^i} &= \frac{m}{2}\, \ddot{\bar{h}}_{ij}(\mathbf{0},t) \, \xi^j .\label{HamiltonEqs}
\end{align} 
By using the Hamilton's equations (\ref{HamiltonEqs0}) and (\ref{HamiltonEqs}),
and replacing $\bar{h}_{ij}(t,\mathbf{0})$ with its Fourier mode decomposition \eqref{eq:H-lambda},
\begin{equation}
\bar{h}_{ij}(t,\mathbf{0})\,  =\, \frac{2 \kappa}{\ell^{3/2}}\sum_{\lambda, \mathbf{k}}\mathcal{A}_{\lambda,\mathbf{k}}(t) e^{\lambda}_{ij},
\label{variables-new}
\end{equation}
we obtain the geodesic deviation equation as
\begin{align}
\ddot{\xi}^i = \frac{\kappa}{\ell^{3/2}} \sum_{\lambda}\sum_{\mathbf{k}} \ddot{\mathcal{A}}_{\lambda, \mathbf{k}}\, e^{\lambda}_{ij} \, \xi^j.
\label{arm-EoM}
\end{align}
This equation represents the interaction between the detector and the (effective) perturbed metric. Moreover,  it gives the tidal acceleration of $\xi$ in the presence of GWs. For each mode ${\bf k}$, the equations of motion become
\begin{subequations}
\begin{align}
\ddot{\xi}^1_{\bf k} &= \frac{\kappa}{\ell^{3/2}} \left[  \ddot{\mathcal{A}}_{+, \mathbf{k}}\,  \xi^1_{\bf k} +  \ddot{\mathcal{A}}_{\times, \mathbf{k}} \, \xi^2_{\bf k}  \right], \label{eq:arm1} \\
\ddot{\xi}^2_{\bf k} &= \frac{\kappa}{\ell^{3/2}} \left[  - \ddot{\mathcal{A}}_{+, \mathbf{k}}\,  \xi^2_{\bf k} +  \ddot{\mathcal{A}}_{\times, \mathbf{k}} \, \xi^1_{\bf k}  \right]. \label{eq:arm2}
\end{align} \label{eq:arm-tot}
\end{subequations}
In the rest of this section, our aim will be to analyze the solutions of $\ddot{\xi}^1_{\bf k}$ and $\ddot{\xi}^2_{\bf k}$ when  the behaviour of the strain $\mathcal{A}_{\lambda, \mathbf{k}}$ is known. To be more precise, it is provided by the solutions of the $\mathcal{A}_{\lambda, \mathbf{k}}$, given by the effective evolution equations of the polymer effective Hamiltonians  (\ref{PolymerE}) or (\ref{PolymerA}).

As mentioned in subsection \ref{PolySubSection}, polymer representations usually come into two polarizations:
either $\hat{\mathcal{A}}$ is not well-defined but $\hat{\mathcal{E}}$ is, or vice versa.
In the polarization where $\hat{\mathcal{A}}$ is not well-defined, the spectrum
of its conjugate variable $\hat{\mathcal{E}}$ becomes discrete. This is basically because there is no $\hat{\mathcal{A}}$ on the Hilbert space to generate infinitesimal transformations in $\hat{\mathcal{E}}$. The inverse of this statement is valid for the case where $\hat{\mathcal{E}}$ is not well-defined. We will consider both cases in what follows. However, note that in LQG, the connection is holonomized/polymerized and the triad is discretized. In our notation, $\mathcal{A}$ corresponds to the metric perturbations [see Eq.~\eqref{def-q}], hence a polarization where ${\cal E}_{\mathbf{\lambda,k}}$ is polymerized, resulting in  ${\cal A}_{\lambda,\mathbf{k}}$ becoming discrete, is more in line with LQG \cite{Garcia-Chung:2020zyq}. This is the case that we are more interested about.


\subsection{Perturbative analysis}
\label{solutions:perturbative}

Assuming initial length  $\xi_0$ for the detector's arm, we can set $\xi(t) = \xi_0 + \delta \xi(t)$, in which $\delta \xi(t)$ is the displacement induced by the GWs. This, applied to each mode and each polarization, yields
\begin{subequations}
\begin{align}
\xi^1_{\bf k}(t) &= \xi^1_0 +\delta \xi^1_{\bf k}(t), \label{rel:xi1}\\
\xi^2_{\bf k}(t) &= \xi^2_0 +\delta \xi^2_{\bf k}(t). \label{rel:xi2} 
\end{align}\label{rel:xi-tot}
\end{subequations}
Using Eqs.~\eqref{eq:arm-tot} to the leading order, we obtain the following equations of motion for the arm's displacements,
\begin{subequations}
\begin{align}
\delta \ddot{\xi}^1_{\bf k} &=  \frac{\kappa}{\ell^{3/2}} \left[  \ddot{\mathcal{A}}_{+, \mathbf{k}}\,  \xi^1_0 +  \ddot{\mathcal{A}}_{\times, \mathbf{k}} \, \xi^2_0  \right], \\
\delta \ddot{\xi}^2_{\bf k} &=  \frac{\kappa}{\ell^{3/2}} \left[ -  \ddot{\mathcal{A}}_{+, \mathbf{k}}\,  \xi^2_0 +  \ddot{\mathcal{A}}_{\times, \mathbf{k}} \, \xi^1_0  \right],
\end{align}
\end{subequations}
where $\delta \xi^{1, 2}_{\mathbf{k}}$ and $\mathcal{A}_{\lambda, \mathbf{k}}$ (with $\lambda=+,\times$) are considered as small perturbations. We now integrate these equations to obtain the general solutions
\begin{subequations}
\begin{align}
\delta {\xi}^1_{\bf k}(t) &=  \frac{\kappa}{\ell^{3/2}} \left[  {\mathcal{A}}_{+, \mathbf{k}}(t)\,  \xi^1_0 +  {\mathcal{A}}_{\times, \mathbf{k}}(t) \, \xi^2_0  \right] + v^1_0 \, t, \label{rel:deltaxi1}\\
\delta {\xi}^2_{\bf k}(t) &=  \frac{\kappa}{\ell^{3/2}} \left[ -  {\mathcal{A}}_{+, \mathbf{k}}(t)\,  \xi^2_0 +  {\mathcal{A}}_{\times, \mathbf{k}}(t) \, \xi^1_0  \right] + {v}^2_0 \, t , \label{rel:deltaxi2}
\end{align} \label{rel:deltaxi-tot}
\end{subequations}
where ${v}^1_0$ and ${v}^2_0$ are integration constants. Note that in this  configuration, one detector's arm is initially at the origin and the other at the position $(\xi^1_0, \xi^2_0)=(\xi_0 \cos\theta, \xi_0 \sin\theta)$. Therefore, GWs propagating in the direction perpendicular  to the $\xi^1-\xi^2$ plane, induce the displacements 
\begin{subequations}
\begin{align}
{\xi}^1_{\bf k}(t) &=  \frac{\kappa}{\ell^{3/2}} \left[  {\mathcal{A}}_{+, \mathbf{k}}(t)\,  \xi_0 \cos\theta +  {\mathcal{A}}_{\times, \mathbf{k}}(t) \, \xi_0 \sin\theta  \right] +  \xi_0 \cos\theta, \label{detector-pert1} \\
{\xi}^2_{\bf k}(t) &=  \frac{\kappa}{\ell^{3/2}} \left[ -  {\mathcal{A}}_{+, \mathbf{k}}(t)\, \xi_0 \sin\theta +  {\mathcal{A}}_{\times, \mathbf{k}}(t) \, \xi_0 \cos\theta  \right] + \xi_0 \sin\theta , \label{detector-pert1}
\end{align} \label{detector-pert-tot}
\end{subequations}
on the detector's arm. Let us now assume  that the metric perturbation  has only a ${\mathcal{A}}_{+, \mathbf{k}}(t)$ polarization. Then the solutions (\ref{detector-pert-tot}) reduce to
\begin{subequations}
\begin{align}
{\xi}^1_{\bf k}(t) &=  \left[1+ \frac{\kappa}{\ell^{3/2}} {\mathcal{A}}_{+, \mathbf{k}}(t) \right] \xi_0 \cos\theta, \label{rel:xi1Plus} \\
{\xi}^2_{\bf k}(t) &=  \left[1 - \frac{\kappa}{\ell^{3/2}} {\mathcal{A}}_{+, \mathbf{k}}(t) \right] \xi_0 \sin\theta .\label{rel:xi2Plus}
\end{align} \label{rel:xiPlus-tot}
\end{subequations}
To study the effects of the polymer quantum dynamics  on the arm's length, we will substitute the effective  solutions of GWs [cf. Eqs.~(\ref{eq:EoM-eff-E-sol-app1}) and (\ref{eq:EoM-eff-A-sol-app})] into Eqs.~\eqref{rel:xiPlus-tot}  to obtain the evolution of detector's arms
\begin{subequations}
\begin{align}
{\xi}^1_{\bf k}(t) &=  \left[1+ \frac{1}{2} \bar{h}^{(\mathcal{E} / \mathcal{A})}_{+,\mathbf{k}}(t) \right] \xi_0 \cos\theta, \label{rel:xi1Plus1H} \\
{\xi}^2_{\bf k}(t) &=  \left[1 - \frac{1}{2} \bar{h}^{(\mathcal{E} / \mathcal{A})}_{+,\mathbf{k}}(t) \right] \xi_0 \sin\theta ,\label{rel:xi2Plus2H}
\end{align}\label{rel:xiPlus-tot2}
\end{subequations}
and then compare the geometry of the arm's displacements induced by the plus polarization of GWs, with those given by the classical GR. 

\begin{figure}
	\begin{subfigure}[b]{0.45\textwidth}
		\includegraphics[width=1.0\textwidth]{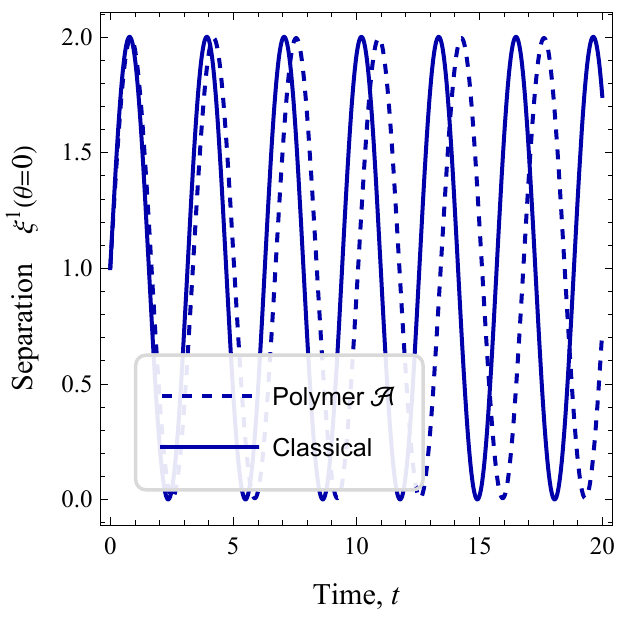}
		\caption{}
		\label{fig:GWPlusPhasesA}
	\end{subfigure}
	\begin{subfigure}[b]{0.45\textwidth}
		\includegraphics[width=1.0\textwidth]{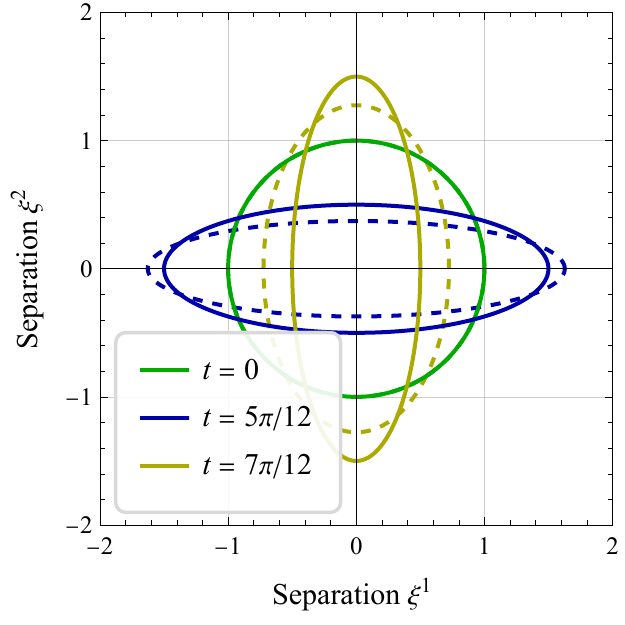}
		\caption{}
		\label{fig:GWPlusPhasesDiffA}
	\end{subfigure} 
	\caption{The time-evolution of the `plus polarization' mode of GWs for the classical (solid curves) and the polymer $\cal{A}$ (dashed curves) solutions. Plots are obtained with the choice of parameters $\bar{h}_I = \xi_0 = \bar{\nu} = c = \hbar = 1, k = 2$. The case (a) shows the detector's arm behavior due to both classical and polymer GWs, and the case (b) represents a comparison between the classical and polymer solutions at three different times.}\label{fig:GWPlusPolyA}
\end{figure}

The displacements in detector's arms, corresponding to the plus polarization of GWs,  under the effect of the classical and (the case $\mathcal{A}$) polymer GWs  [cf. Eqs.\eqref{rel:xiPlus-tot2} together with the solution (\ref{eq:EoM-eff-A-sol-app2})], are depicted in Fig.~\ref{fig:GWPlusPolyA}.
Fig.~\ref{fig:GWPlusPhasesA} represents the time evolution of the displacements in classical and polymer scenarios, whereas Fig.~\ref{fig:GWPlusPhasesDiffA} shows a comparison between the classical  and polymer  cases   in three instances of time. Note that for the classical solution we used the expression
\[\bar{h}^{\rm class}_{+,\mathbf{k}}(t) = \bar{h}_I \cos\left(kc t \right).\]

\begin{figure}
	\centering
	\begin{subfigure}[b]{0.45\textwidth}
		\includegraphics[width=1.0\textwidth]{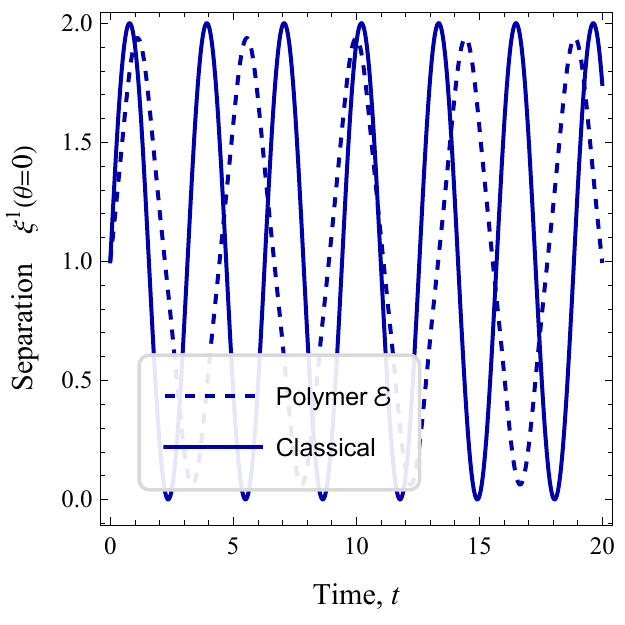}
		\caption{}
		\label{fig:GWPlusPhasesE}
	\end{subfigure}
	\begin{subfigure}[b]{0.45\textwidth}
		\includegraphics[width=1.0\textwidth]{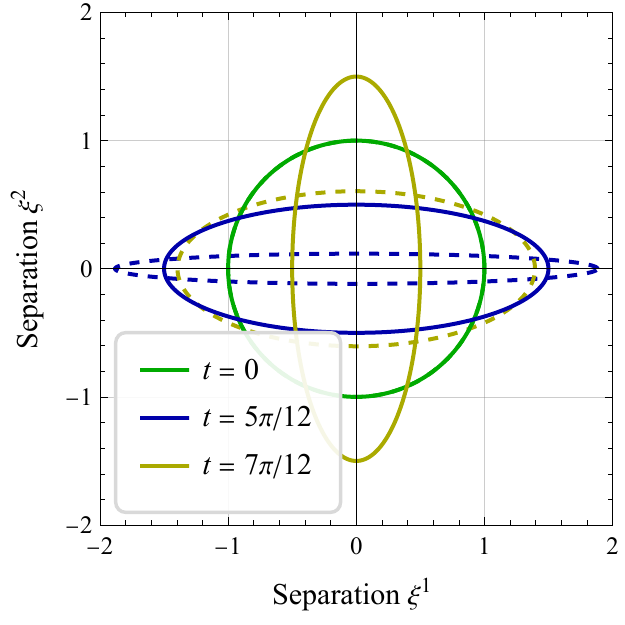}
		\caption{}
		\label{fig:GWPlusPhasesDiffE}
	\end{subfigure}
	\caption{The time-evolution of the plus polarization mode of GWs for classical (solid curves) and polymer $\cal{E}$ (dashed curves) solutions. Plots are obtained with parameters $\bar{h}_I = \xi_0 = \bar{\nu} = c = \hbar = 1, k = 2$. In (a) the detectors arm behavior due to both classical and polymer gravitational waves are shown, and in (b) a comparison between the classical and polymer solutions at three different times is made.}\label{fig:GWPlusPolyE}
\end{figure}

Likewise, Fig.~\ref{fig:GWPlusPolyE} depicts the behaviours of the displacements in detector's arm ensued from the  plus polarization of the polymer $\cal{E}$ solution [cf. Eq.~\eqref{eq:EoM-eff-E-sol-app2}]. A comparison with the classical case is also made.  It is interesting  to mention that the polymer corrections in solution \eqref{eq:EoM-eff-E-sol-app2} depend on the mode of the GW, which means that different modes of a GW induce different displacements on the detector's arms. This phenomenological property might have observational consequences, as will be discussed at the end of this subsection and in Sec.~\ref{sec:LISA}. In Fig.~\ref{fig:GWPlusPhasesDiffE} we can see that the case $\cal{E}$ polymer corrections produce larger effects compared to the polymer $\cal{A}$ corrections at the same three time instances with the same control parameters (cf. Fig.\ref{fig:GWPlusPhasesDiffA})

An analysis of polymer corrections in different scenarios is in order. From Eqs.~\eqref{eq:EoM-eff-A-sol-app2} and \eqref{eq:EoM-eff-E-sol-app2} we see that
the amplitude of the GWs attenuates as 
$\delta \bar{h}^{(\mathcal{A})} \sim \bar{h}_I^3 \bar{\nu}^2/\hbar^2$,
in the Polymer $\mathcal{A}$ case, while it attenuates as 
$\delta \bar{h}^{(\mathcal{E})}  \sim \bar{h}_I^3 k^2 \bar{\mu}^2/\hbar^2$,
in the Polymer $\mathcal{E}$ case.
Moreover, if $\bar{\mu} \sim \bar{\nu}$, the attenuation in perturbation $\bar{h}_{+,\mathbf{k}}$ in polymer $\mathcal{E}$ case is about $3\times k^2$ times larger than that in the polymer $\mathcal{A}$ case.
On the other hand, by assuming that in a binary system $\bar{h}_I \sim 10^{-23}$ \cite{Flanagan:2005yc}, we find that  the corrections in the amplitude of the GWs due to polymer effects are around $\delta \bar{h}_I^{(\mathcal{A})} \sim \, \bar{\nu}^2$ (in the Polymer $\mathcal{A}$ case) and $\delta \bar{h}_I^{(\mathcal{E})} \sim \, k^2\bar{\mu}^2$ (in the Polymer $\mathcal{E}$ case).

The high energy sources can emit GWs with frequencies  around $f \sim 10^4$Hz \cite{Flanagan:2005yc}, thus the GWs with short wavelength in Polymer $\mathcal{E}$ case would decay faster than those with long wavelength.
In the present setting, the length of the detector's arm, $\xi_0$, can amplify tiny amplitudes of GWs. In today's technology, there is a limitation for the arm's length of the GW detectors; in high frequencies $1 \lesssim f \lesssim 10^4$ Hz, it is about $3\, {\rm km}\lesssim \xi_0 \lesssim 4\,{\rm km}$  (see \cite{LIGOScientific:2014pky, VIRGO:2014yos, KAGRA:2018plz} for LIGO, VIRGO and KAGRA collaborations, respectively). 
Therefore, it would be better to investigate lower frequencies with larger arm's lengths where GWs have stronger amplitudes. The mission of the LISA space probe is  to detect and measure GWs  produced by mergers of supermassive black holes \cite{amaro2017laser}, so LISA might be a suitable candidate for the detection of the tiny effects of the polymer corrections in the gravitational waveforms. We will investigate  the imprints of the polymer quantization schemes in LISA in the Sec.~\ref{sec:LISA}, however, it is instructive to do an order analysis of the polymer corrections in the lower frequency ranges. If the amplitude of GW is of order $\bar{h}_I \sim 10^{-15}$, then attenuation coming from polymer quantization in case $\mathcal{E}$ will be at the order of $\delta \bar{h}^{(\mathcal{E})}_I \sim 10^{-45} \, \left(\bar{\mu}/\hbar \right)^2$ for $k \sim 1$. In this case, if we assume $\bar{\mu} \sim 10^{10}\hbar$ (remember that in principle, the rescaled parameter $\bar{\mu}$ can be larger than the bare polymer scale $\mu$; see \eqref{eq:EoM-eff-E-sol-app2} for more details), the attenuation will be $\delta \bar{h}^{(\mathcal{E})}_I \sim 10^{-25}$, which is not far from the current observational capability. Exact numerical details in this regard will be presented in Sec.~\ref{sec:LISA}.


\subsection{Numerical analysis}

In this section we examine numerical (i.e., non-perturbative) solutions to the polymer and classical equations of motion in order to validate the treatment presented in the previous section. Here we directly integrate quations \eqref{eq:arm1} and \eqref{eq:arm2} numerically for the same choice of parameters and conditions considered in Sec.~\ref{sec:phenomenology}, for the polymer $\mathcal{E}$ case. We additionally choose the polymer scale $\mu$ and GW amplitude to be unrealistically large so to qualitatively demonstrate the impact of nonlinear effects.

We will restrict our analysis to the case where the GW amplitude $\bar{h}_I$ is small, $\mathcal{O}(10^{-8})$, so that nonlinear gravitational contributions to the behavior should be below the level of numerical roundoff, although note that we will still consider a numerical solution to the full equations of motion, \eqref{eq:arm1} and \eqref{eq:arm2}. This nevertheless reduces any terms quadratic in the amplitude to effectively zero. We then consider the polymer scale to be large, as to qualitatively demonstrate polymer effects on the arm behavior, although in practice we do not expect such large polymer scale values.

\begin{figure}[htb]
	\centering
	\begin{subfigure}[b]{3in}
		\includegraphics[width=2.6in]{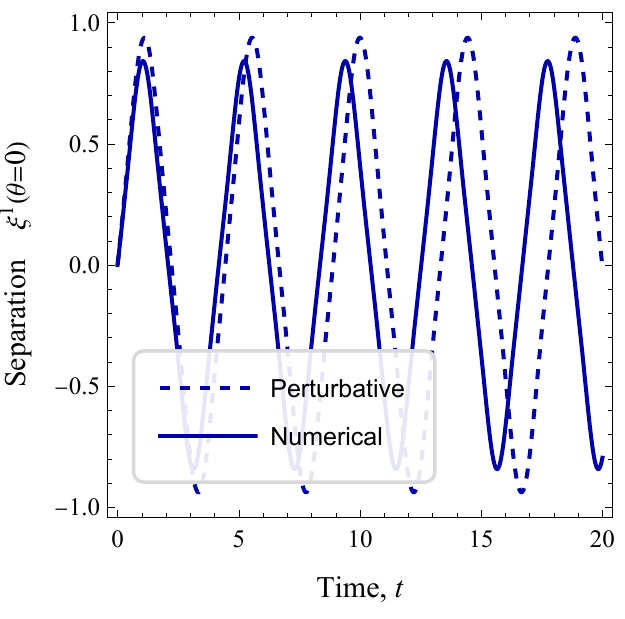}
		\caption{}
		\label{fig:SepNumComp}
	\end{subfigure}
	\begin{subfigure}[b]{3in}
		\includegraphics[width=2.6in]{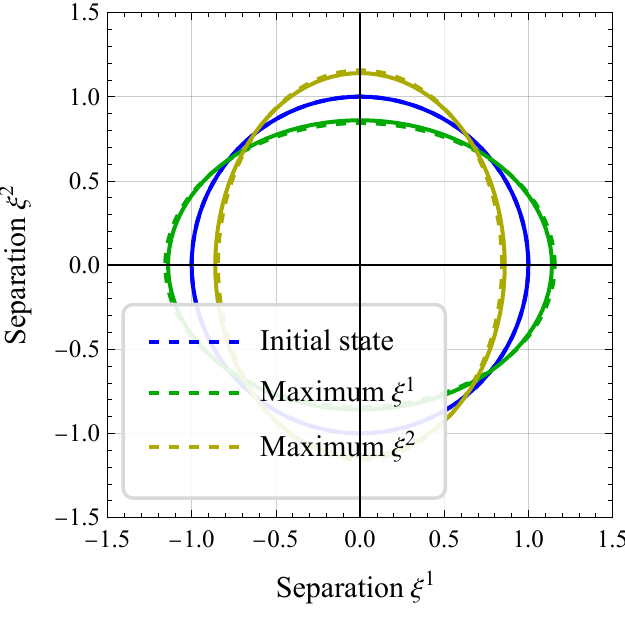}
		\caption{}
		\label{fig:GWPlusPolymer}
	\end{subfigure}
	\caption{Left: GW arm separation according to the perturbative polymer and full numerical solutions, for a large (but still strongly perturbative) GW amplitude $\bar{h}_I = 10^{-8}$. An extremely large value for $\mu = \bar{h}_I^{-1}$ is chosen to exaggerate nonlinear polymer effects, which are apparent. For smaller $\mu$, the perturbative and numerical solutions agree. Right: Polarization of the GWs according to the perturbative and full numerical solutions, again exaggerated to demonstrate the difference between the solutions.
	}\label{fig:GWPlus}
\end{figure}

As noted in our previous work \citep{Garcia-Chung:2020zyq,Garcia-Chung:2021doi}, polymer GWs  will undergo both a frequency and amplitude shift relative to the classical solution. Over large distances, this can appear as an order unity phase shift in the GWs. The linearized equations of motion suggest the detector arm will directly probe the GW (e.g. Eqs.~\eqref{rel:xi1Plus} and \eqref{rel:xi2Plus}), with only a nonlinear coupling term that will be extremely small.

As suggested by the form of Eqs.~\eqref{eq:arm1} and \eqref{eq:arm2}, the solution for the arm separation is directly related to that of the gravitational waveform itself when the amplitudes involved are small. We see precisely this behavior in Figure \ref{fig:SepNumComp}. Note that this is for a very large polymer scale where nonlinear corrections are important; as the polymer scale (and GW amplitude) become smaller as expected for observations of GWs, the two solutions are found to agree, and the perturbative picture is recovered.


\section{How loud polymer effects will be in LISA\label{sec:LISA}}

The Laser Interferometer Space Antenna (LISA) is a proposed space probe for GW signals and will open the mHz band for exploration of GWs. Sensitivity curves can be used for surveying the types of gravitational systems that can be observed by the LISA
 mission \cite{amaro2017laser, Moore:2014lga}. Here we use the sensitivity curve to explore the polymer effects in LISA detectors. We will compute the signal to noise ratio for simple binary systems and calculate the order of polymer corrections. We consider sky-averaged sensitivities using the latest quantities and methods described for design parameters in Refs.~\cite{Larson:1999we, Cornish:2001bb}.
 
In the frequency domain, the strain induced by the amplitude of GWs can be written as
\begin{equation}
\bar{h}_{\mathbf{k}}(f) = {\cal R}^+(f)\, \bar{h}_{+, \mathbf{k}}(f) + {\cal R}^\times(f)\, \bar{h}_{\times, \mathbf{k}}(f) \, , \label{rel:strain}
\end{equation}
in which ${\cal R}^+(f)$  and ${\cal R}^\times(f)$ are the detector response functions for each polarization, and $\bar{h}_{+, \mathbf{k}}$, $\bar{h}_{\times, \mathbf{k}}$ are given by Eqs.~\eqref{eq:EoM-eff-E-sol-app2} and \eqref{eq:EoM-eff-A-sol-app2} for each scheme of polymerization, $\mathcal{E}$ or $\mathcal{A}$. The averaged magnitude of the spectral power of the signal in the detector, $\left\langle \bar{h}_{\mathbf{k}}(f) \bar{h}_{\mathbf{k}}^*(f) \right\rangle$, is related to the magnitude of the spectral power of each polarization, $|\bar{h}_{+, \mathbf{k}}(f)|^2$ and $|\bar{h}_{\times, \mathbf{k}}(f)|^2$, as
\begin{eqnarray}\label{averagedh}
\left\langle \bar{h}_{\mathbf{k}}(f) \bar{h}_{\mathbf{k}}^*(f) \right\rangle = {\cal R}(f)\left(|\bar{h}_{+, \mathbf{k}}(f)|^2 +|\bar{h}_{\times, \mathbf{k}}(f)|^2\right),
\end{eqnarray} 
where, ${\cal R}(f)$ is the averaged detector response function. The majority of sources that LISA can detect are binary systems with different mass ratios. For simplicity, we consider spinless binary systems with comparable masses \cite{Cornish:2017vip}. In order to plot dimensionless characteristic strain, given by $h_c(f) = \sqrt{f\;S(f)}$ where $S(f) = 16/5 f \bar{h}_{\mathbf{k}}^2(f)$, we need first to calculate $S(f)$ using the amplitude of the wave, $\bar{h}_{\mathbf{k}}(f)$, in frequency domain (given by Eq.~\eqref{rel:strain}). The orbit of the binary system might have inclination relative to the line of sight, the factor $16/5$ comes from averaging over the inclination and polarizations of GWs \cite{Robson:2018ifk}. 

The coalescence of two black holes has three stages; inspiral (post-Newtonian regime), merger (relativistic regime) and ring-down (relativistic perturbative regime) phases. We only have a theoretical model for the inspiral phase, thus we use this model to compute the effects of the polymerization of GWs during the inspiral phase (a phenomenological template can be found in Ref.~\cite{Ajith:2007qp} for each phase of this gravitational source). 
We assume that the source is classical and that it produces GWs with initial amplitude and  template for the evolution of the frequency of the inspiral phase. Later, through propagation, the GW waveform receives corrections  effectively for each polymer quantization scheme from Eqs.~\eqref{eq:EoM-eff-A-sol-app2} and \eqref{eq:EoM-eff-E-sol-app2}. Thus, the initial amplitude will be \cite{Creighton:2011zz}
\begin{eqnarray}
\bar{h}_I &\equiv& 4\; \frac{\left( G{\cal  M}/c^3\right)^{5/3}}{D/c} (\pi f_{\rm merg})^{2/3} \, \left(\frac{f}{f_{\rm merg}}\right)^{2/3} , \label{rel:hI}
\end{eqnarray}
in which
\begin{equation}
{\cal M}= (m_1 m_2)^{3/5}/(m_1 + m_2)^{1/5}\,,
\label{eq:phase1}
\end{equation}
and $D$ is the luminosity distance. The transition frequencies $f_{\rm merg}$ denote the beginning of the merger phase \cite{Robson:2018ifk}. The template for the evolution of the frequency, to the leading order from Newtonian contribution, is \cite{Creighton:2011zz}
\begin{equation}
\dot{f} = \frac{96}{5}\, \left( G{\cal  M}/c^3\right)^{5/3} \pi^{8/3} f_{\rm merg}^{11/3} \, \left(\frac{f}{f_{\rm merg}}\right)^{11/3} . \label{rel:fdot}
\end{equation}
According to the effective solutions \eqref{eq:EoM-eff-A-sol-app2} and \eqref{eq:EoM-eff-E-sol-app2}, the phenomenological waveform for propagating GWs will be,
\begin{equation}
\bar{h}_{\mathbf{k}}(t) = \bar{h}^{\rm eff}_{1} \cos \left(\phi^{\rm eff}(t)\right) +  \bar{h}^{\rm eff}_{2} \cos \left(3\phi^{\rm eff}(t)\right), \label{rel:waveformeff}
\end{equation}
where the phase of the waveform evolves approximately as $\phi^{\rm eff} (t) \simeq 2 \pi(f t + \tfrac{1}{2} \dot{f} t^2 + {\cal O}(t^3))$.
As stated before, we assume that the classical source produces the initial amplitude of the propagating wave and determines the dynamics of its phase and the wave receives polymer corrections through propagation in spacetime. This is because the EoMs for GWs are homogeneous (source-free equations) and only describe the propagation of the waves, not their production. Consequently, this means that we have assumed that waves are produced by distance sources, and when they travel through quantized spacetime, their dynamics receives corrections. Thus we use \eqref{rel:hI} as the initial amplitude and \eqref{rel:fdot} as the frequency evolution of the source during the inspiral phase. 
Using these conditions in the solutions \eqref{eq:EoM-eff-E-sol-app2} or \eqref{eq:EoM-eff-A-sol-app2} we compute the polymer corrections in GWs produced by the two massive black hole binaries (MBHBs).

\begin{figure}
	\centering
	\begin{subfigure}[b]{0.45\textwidth}
		\includegraphics[width=0.95\textwidth]{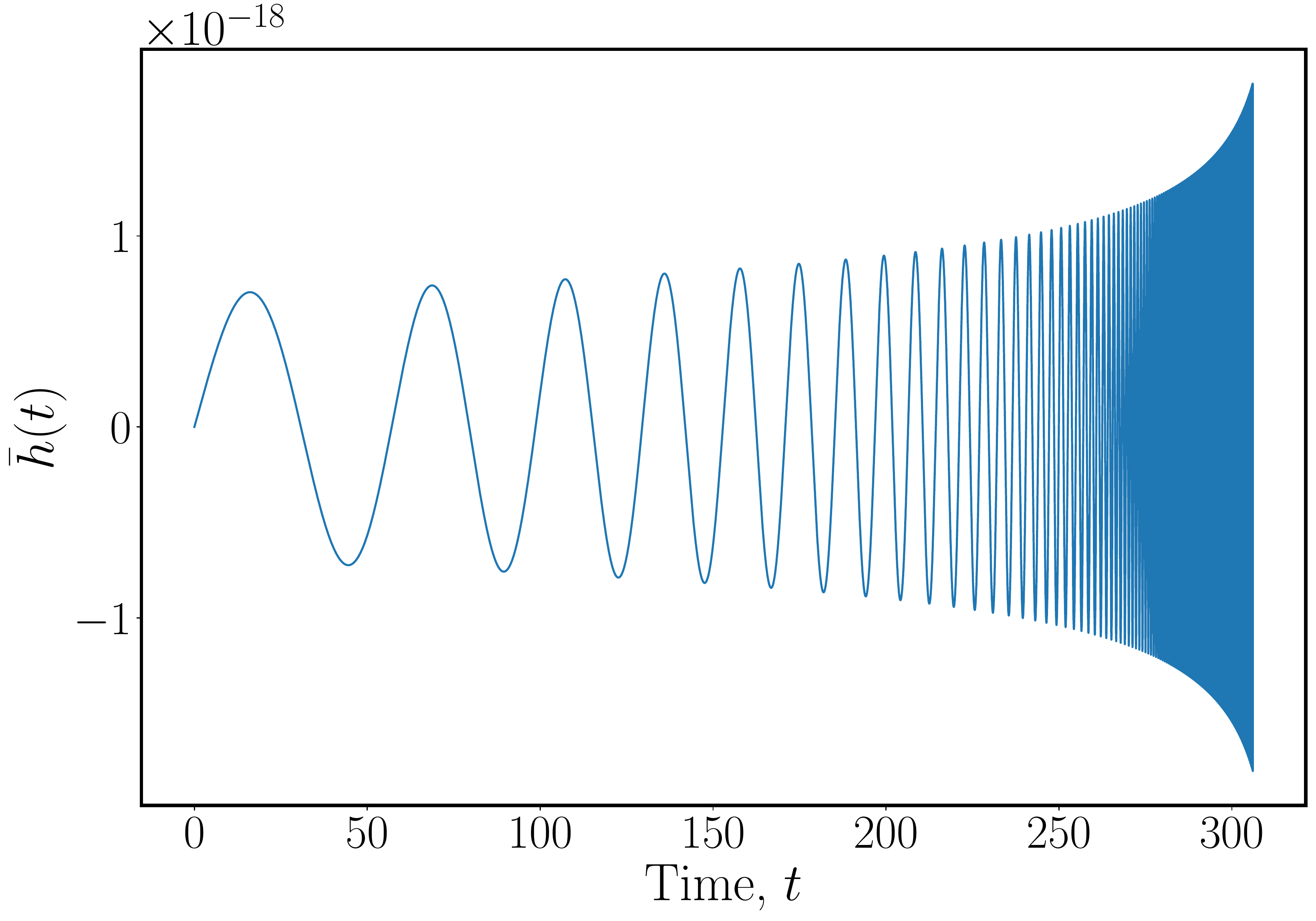}
		\caption{}
		\label{fig:WForm}
	\end{subfigure}
	\begin{subfigure}[b]{0.45\textwidth}
		\includegraphics[width=0.95\textwidth]{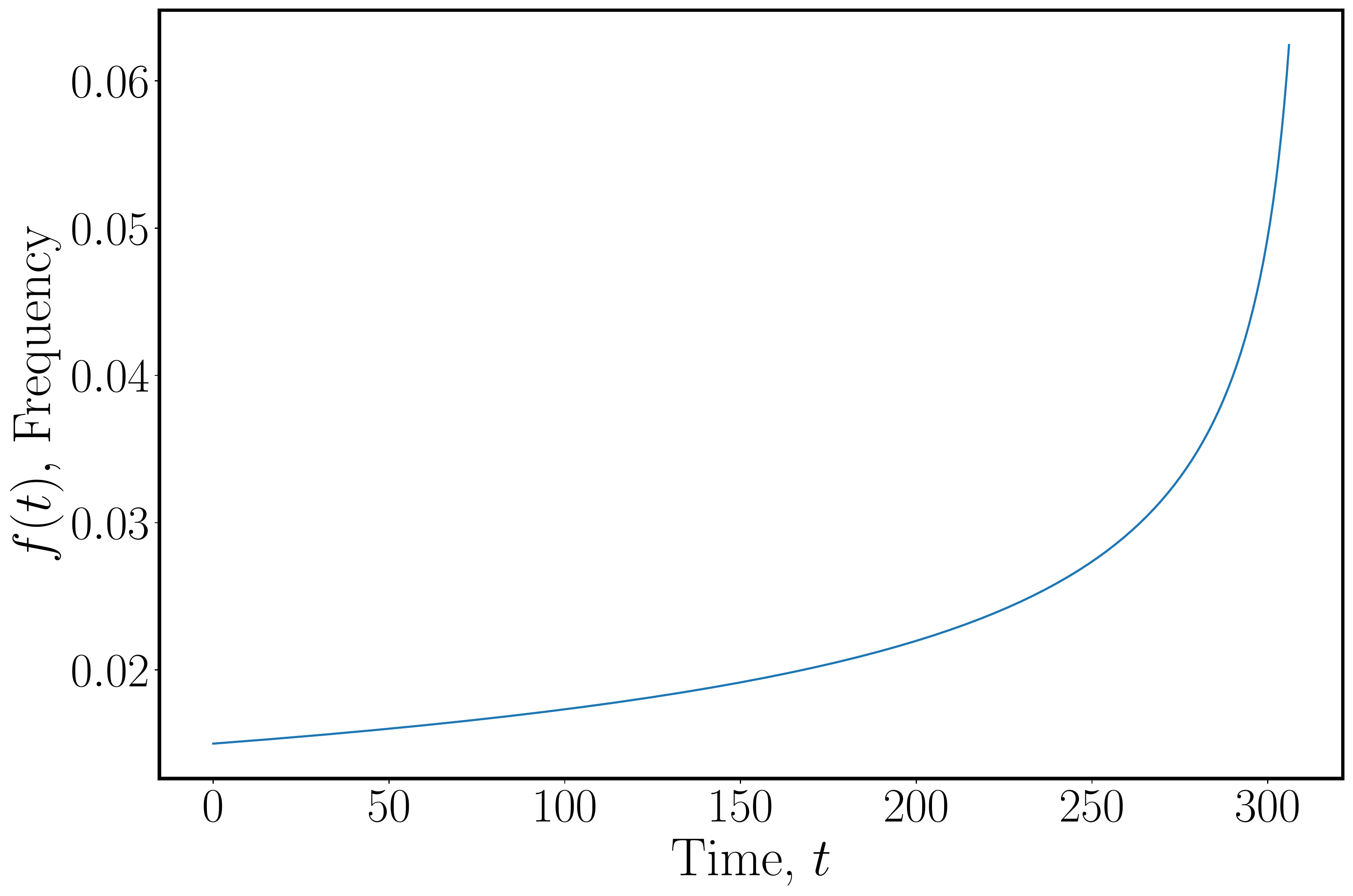}
		\caption{}
		\label{fig:ft}
	\end{subfigure}
	\begin{subfigure}[b]{0.45\textwidth}
	\includegraphics[width=0.95\textwidth]{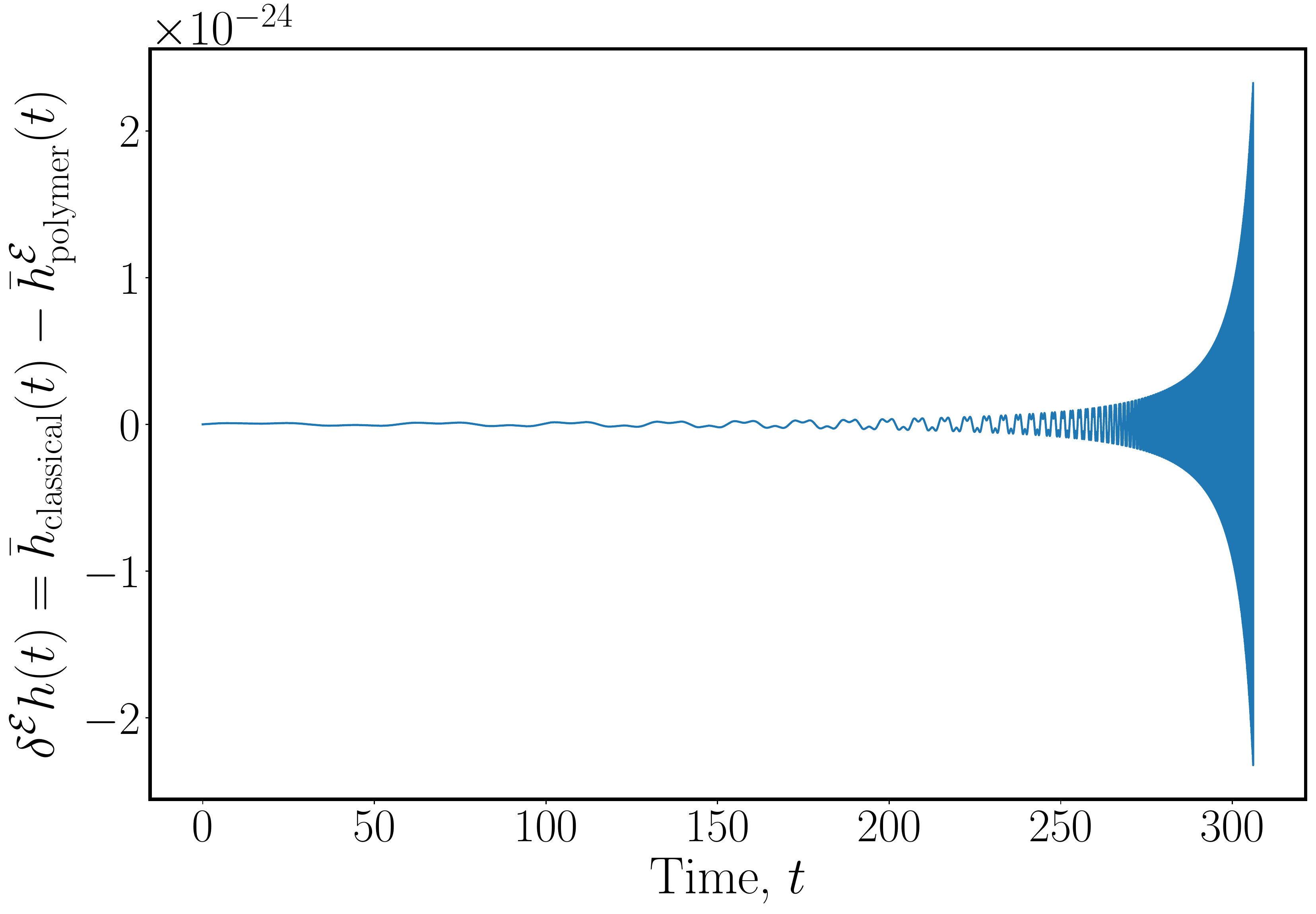}
	\caption{}
	\label{fig:WFPolymerE}
\end{subfigure}
\begin{subfigure}[b]{0.45\textwidth}
	\includegraphics[width=0.95\textwidth]{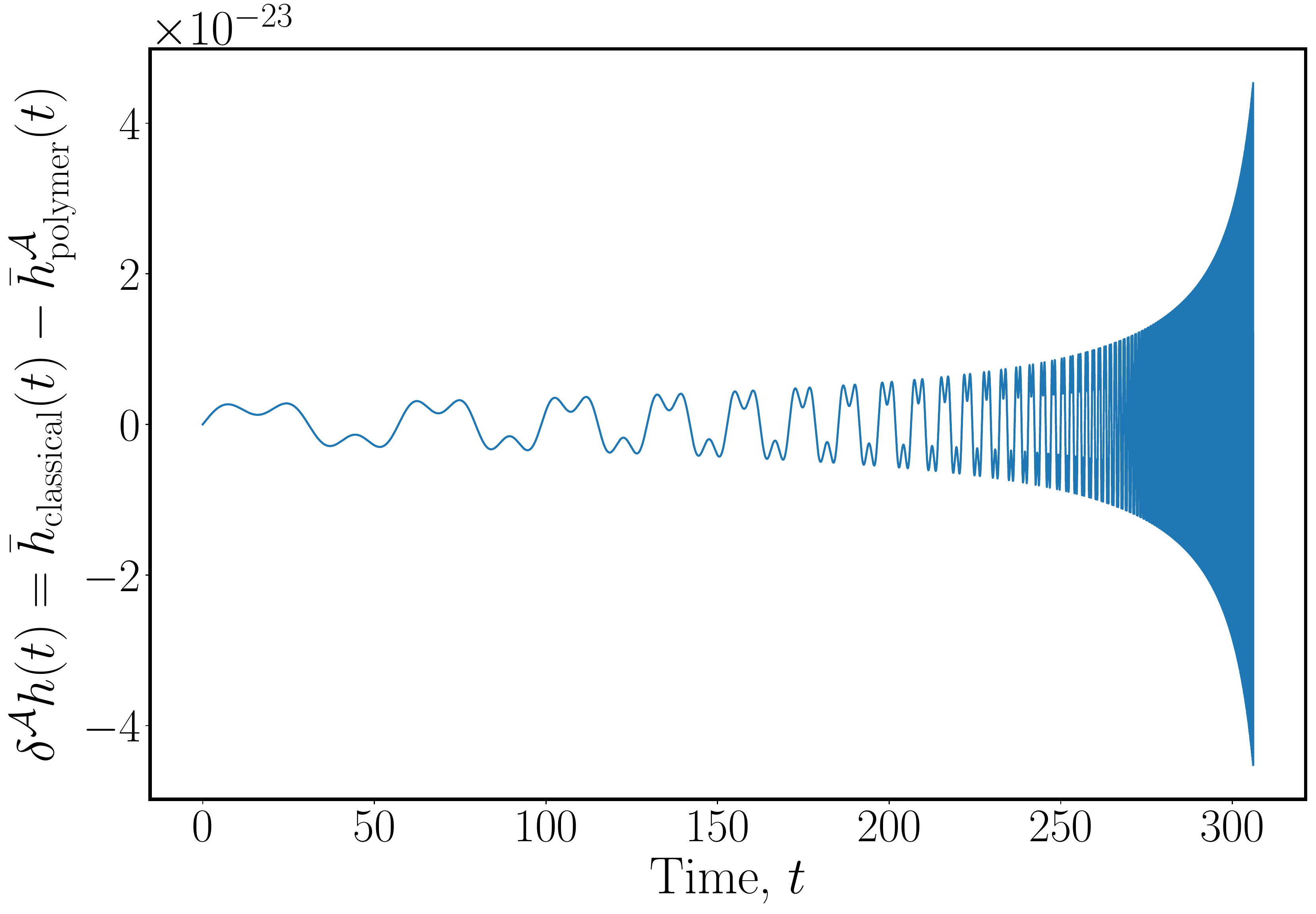}
	\caption{}
	\label{fig:WFPolymerA}
\end{subfigure}
	\begin{subfigure}[b]{0.45\textwidth}
	\includegraphics[width=0.95\textwidth]{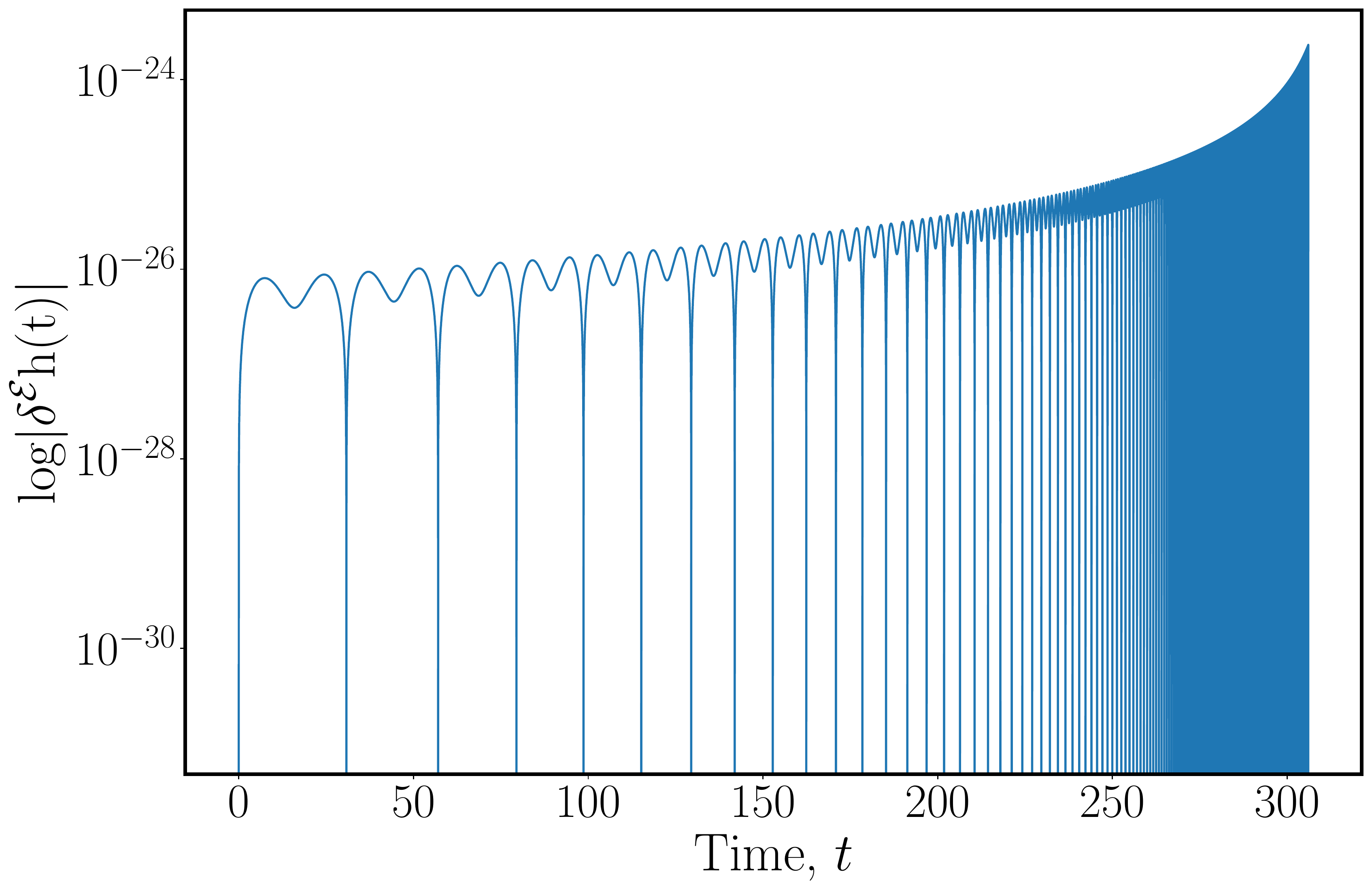}
	\caption{}
	\label{fig:WFPolymerELog}
\end{subfigure}
\begin{subfigure}[b]{0.45\textwidth}
	\includegraphics[width=0.95\textwidth]{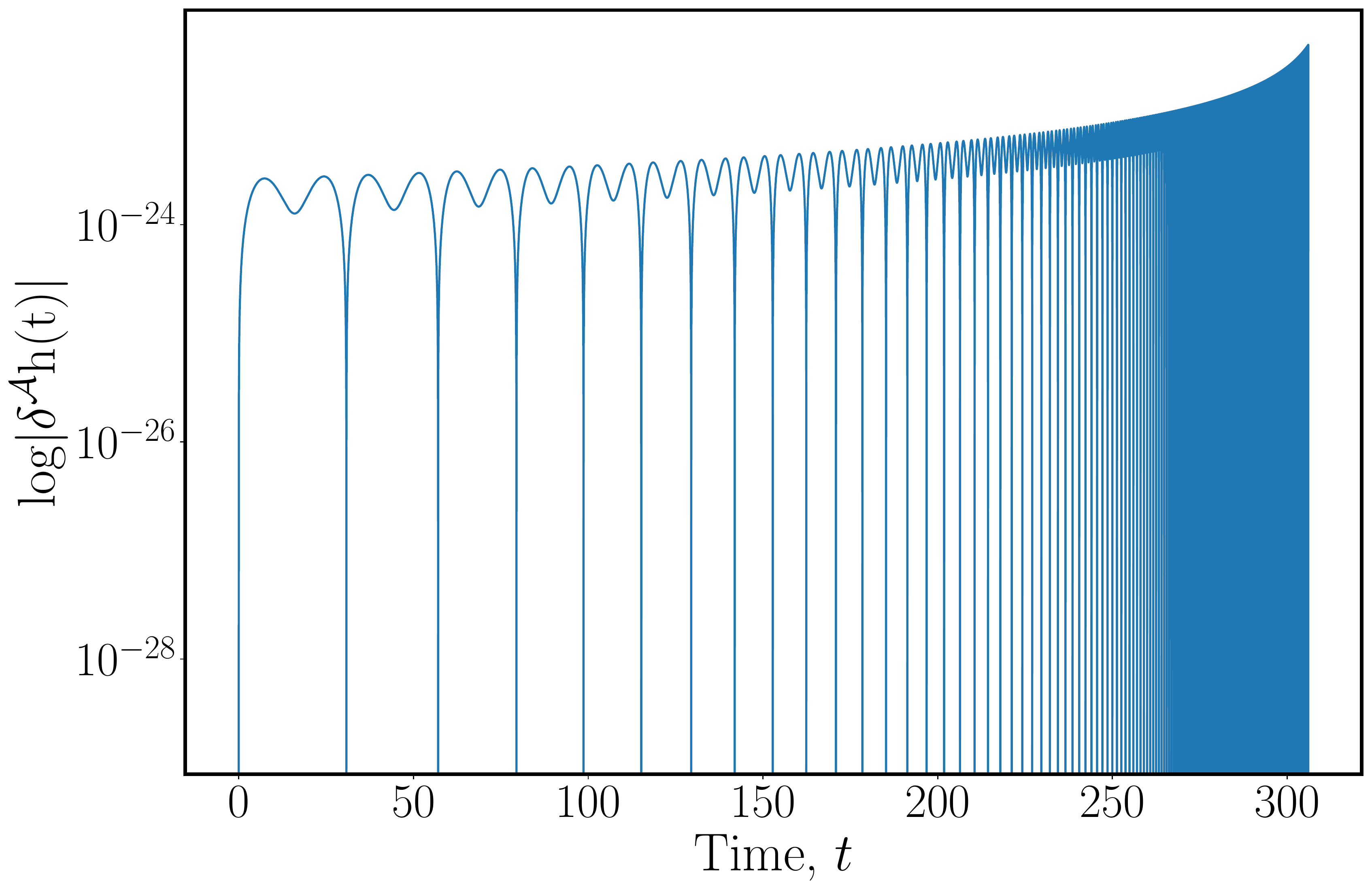}
	\caption{}
	\label{fig:WFPolymerALog}
\end{subfigure}
	\caption{Coalescing waveform of two inspiral black holes with $M \sim 5\times 10^4 M_{\odot}$  at the redshift $z = 0.3$. (a) Waveform shape; (b) evolution of emitted frequency over time; (c) the difference between classical and polymer corrected solution, $\delta h^{\cal E}(t) = h^{\rm class}(t) - h^{\cal E}(t)$ with parameters $\mu = 10^{-50}, \ell = 10^{13} $m; (d) the difference between classical and polymer corrected solution, $\delta h^{\cal A}(t) = h^{\rm class}(t) - h^{\cal A}(t)$ with parameters $\nu = 10^{-58}, \ell = 10^{13}$m; (e) and (f) are the logarithmic plots of the absolute value of the deference functions $\delta h^{\cal E}(t)$ and $\delta h^{\cal A}(t)$}.\label{fig:ht}
\end{figure}

The waveform \eqref{rel:waveformeff} and the expected order of polymer corrections are shown in Fig.~\ref{fig:ht}. We replaced wave number $k$ with frequency $f$ in solution \eqref{eq:EoM-eff-E-sol-app2}, using the dispersion relation \eqref{rel:disperionE}. Figs.~\ref{fig:WForm} and \ref{fig:ft} depict, respectively, the waveform \eqref{rel:waveformeff} and frequency evolution \eqref{rel:fdot}. Figs.~\ref{fig:WFPolymerE} and \ref{fig:WFPolymerA} demonstrate the general behavior of the difference functions $\delta h^{\cal E}(t)$ and $\delta h^{\cal A}(t)$. As the chirp continues, the polymer corrections in polymer $\cal E$ scheme amplify more than polymer corrections in polymer $\cal A$ scheme. This point is specifically illustrated in Figs.~\ref{fig:WFPolymerELog} and \ref{fig:WFPolymerALog}, that the amplitude of the absolute value of difference functions $\delta h^{\cal E}(t)$ and $\delta h^{\cal A}(t)$ are increasing over time and their rate are different for polymer $\mathcal{A}$ and $\mathcal{E}$. To see how loud will be the polymer corrections in LISA, we need to Fourier transform the solutions \eqref{eq:EoM-eff-A-sol-app2} and \eqref{eq:EoM-eff-E-sol-app2}. 
After a few steps we get
\begin{equation}
{\cal F} \left(\bar{h}_{+,\mathbf{k}}(t) \right) = \bar{h}_{+,\mathbf{k}}(f) =  \frac{\bar{h}_I(1-\delta)}{2 \sqrt{\dot{f}}} - \frac{\bar{h}_I \delta}{4 \sqrt{3\dot{f}}}  ,\label{fourierEA}
\end{equation}
where $\delta$ are the corrections  provided by the two polymer schemes, as
\begin{equation}
\delta^{({\cal E})} \equiv \frac{\bar{h}_I ^2 \bar{\mu}^2 k^2}{32 \hbar^2} \quad \text{and} \quad \delta^{({\cal A})} \equiv  \frac{\bar{h}_I ^2 \bar{\nu}^2}{96 \hbar^2}.
\end{equation}
Now we can use Eq.~\eqref{fourierEA} to plot the effective strain spectral density of equal mass black hole inspiral binaries in contrast to the sensitivity curve of LISA (to understand how the sensitivity curve of LISA is calculated see Ref.~\cite{Robson:2018ifk}; here, we have used their expression and repository to calculate the LISA curve).

\begin{figure}
	\begin{subfigure}[b]{0.45\textwidth}
		\includegraphics[width=0.95\textwidth]{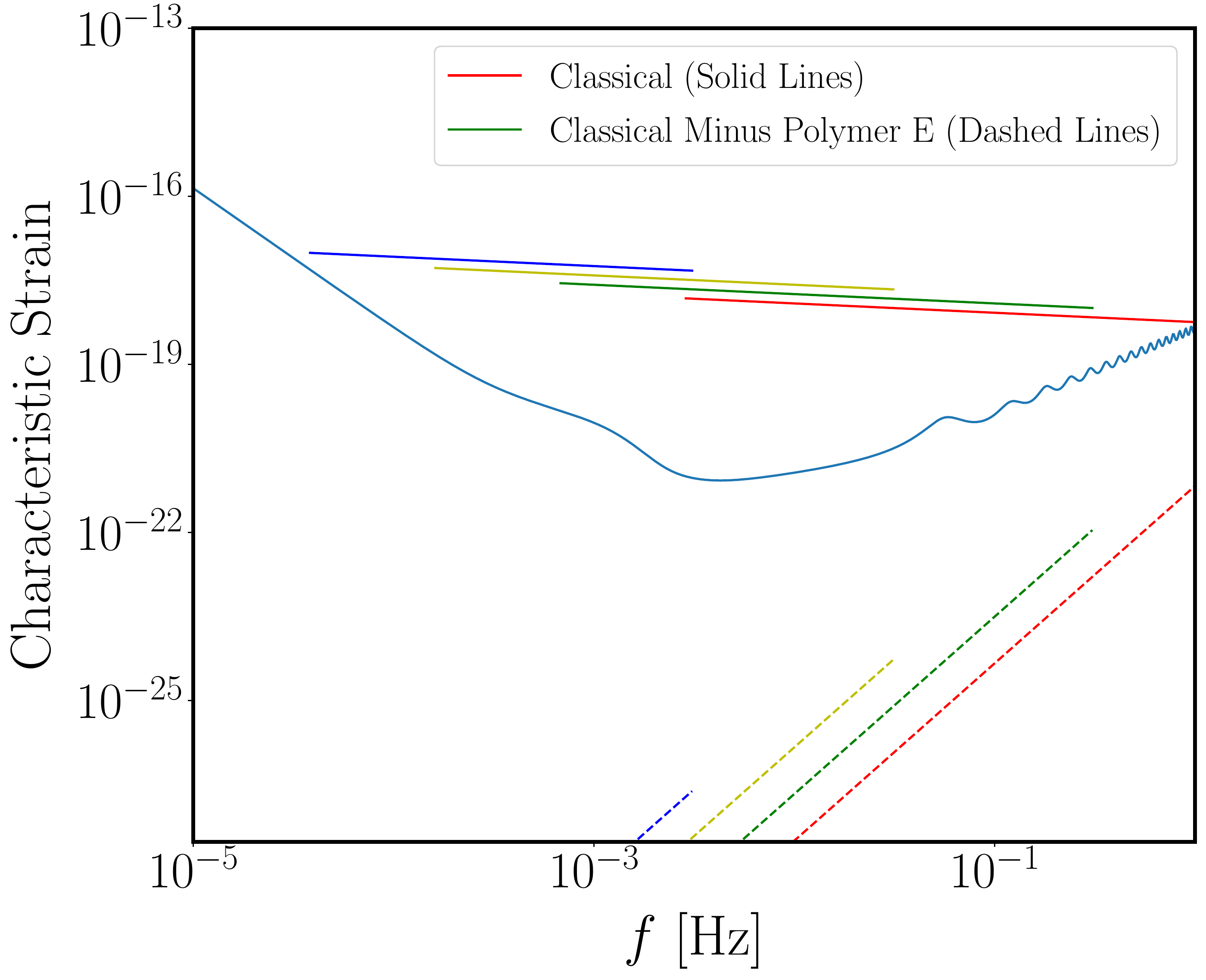}
		\caption{}
		\label{fig:CSE1}
	\end{subfigure}
	\begin{subfigure}[b]{0.45\textwidth}
		\includegraphics[width=0.95\textwidth]{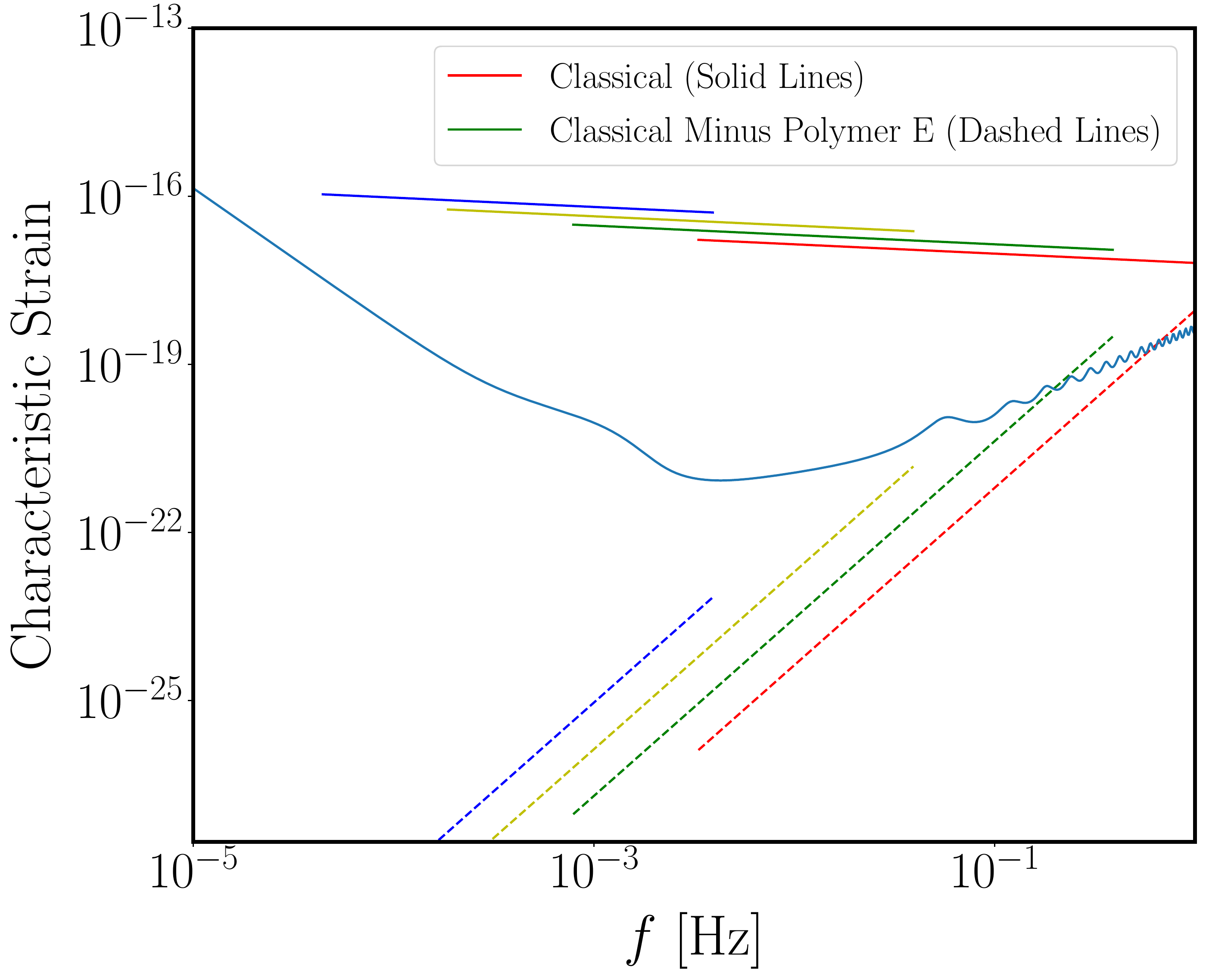}
		\caption{}
		\label{fig:CSE2}
	\end{subfigure}
	\begin{subfigure}[b]{0.45\textwidth}
		\includegraphics[width=0.95\textwidth]{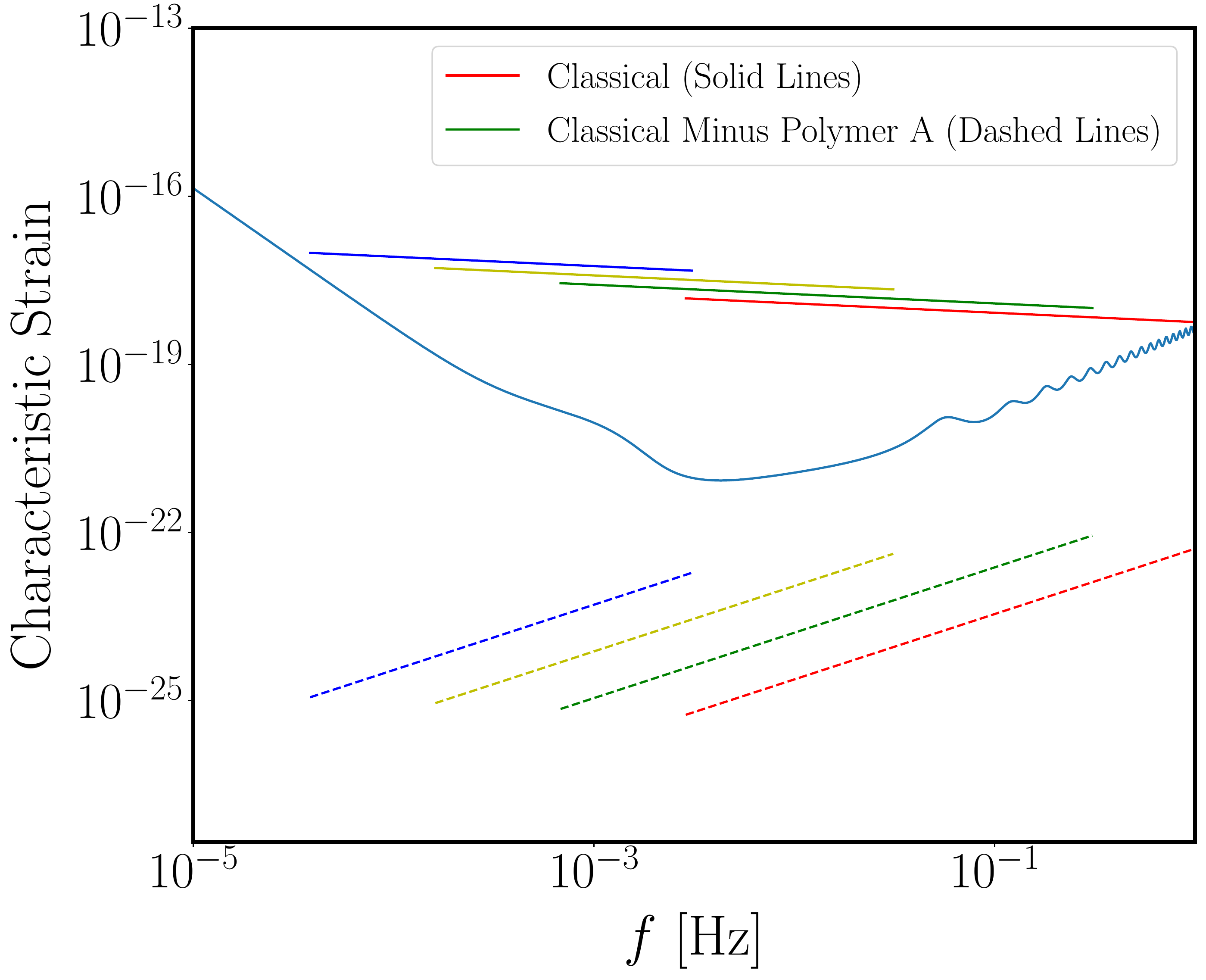}
		\caption{}
		\label{fig:CSA1}
	\end{subfigure}
	\begin{subfigure}[b]{0.45\textwidth}
		\includegraphics[width=0.95\textwidth]{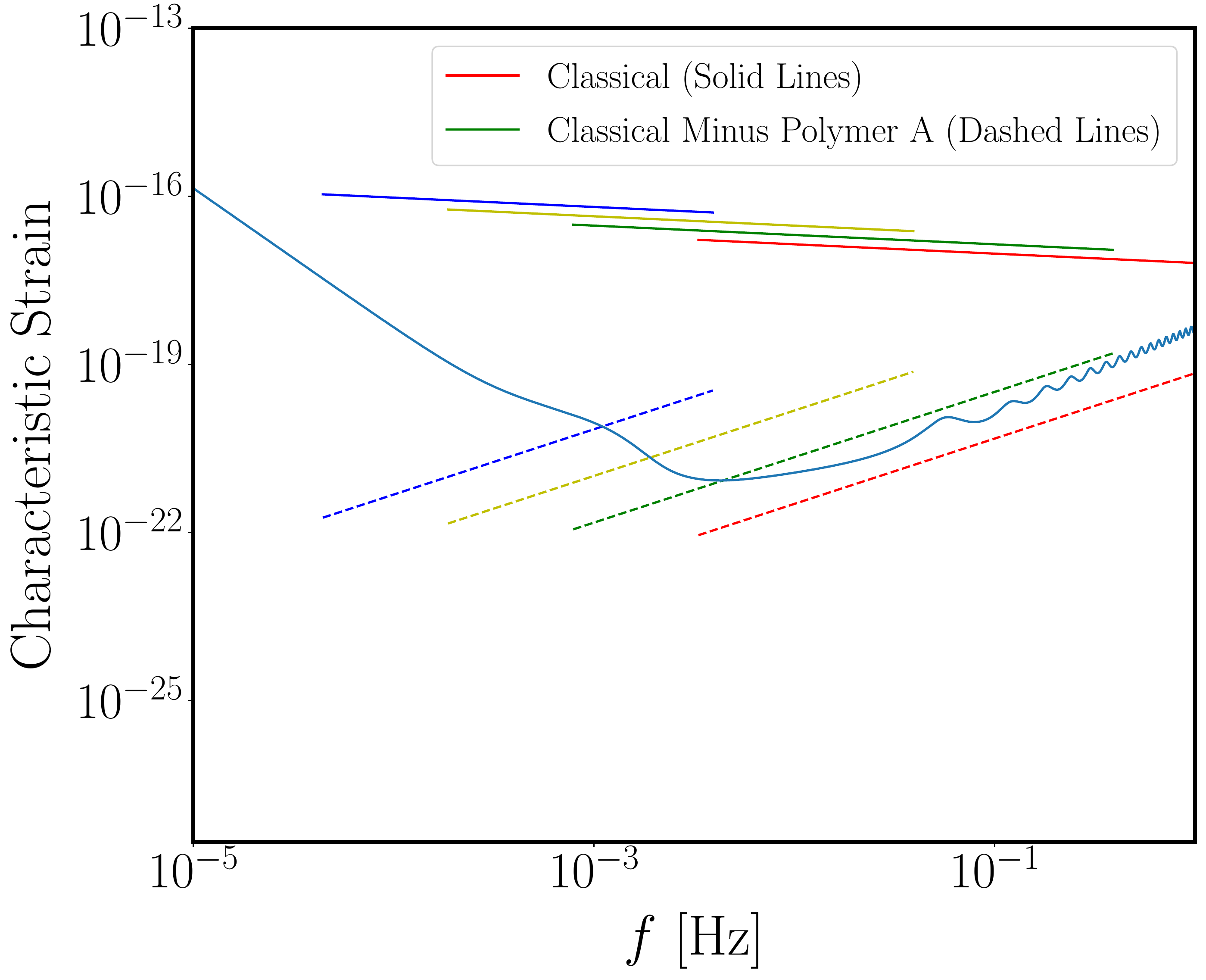}
		\caption{}
		\label{fig:CSA2}
	\end{subfigure}
	\caption{The sensitivity curve in terms of the characteristic strain $\sqrt{f S}$ for four types of signals of inspiral phase of equal mass black hole binaries at $z = 0.3$ (left plots) and $z=0.03$ (right plots), and the source-frame masses $M = 10^3 M_{\odot}~\text{(red lines)}$, $10^4 M_{\odot}~\text{(green lines)}$, $10^5 M_{\odot}~\text{(yellow lines)}$, $10^6 M_{\odot}~\text{(blue lines)}$. Solid lines depict the classical signals $\sqrt{f S_{\rm class}}$ while the dashed lines present the difference between the classical and polymer corrected signals, $\sqrt{f S_{\rm class}}$ - $\sqrt{f S_{\rm poly}}$. Figures  (a) and (b) depict the polymer $\cal E$ scheme with parameters $\mu = 10^{-50}, \ell = 10^{13}$m, while the figures  (c) and (d) depict the polymer $\cal A$ scheme with parameters $\nu = 10^{-58}, \ell = 10^{13}$m}\label{fig:CS}
\end{figure}

Fig.~\ref{fig:CS} shows the characteristic strain and the order of polymer corrections in LISA for four equal mass black hole binaries at two redshifts $z = 0.3$ and $z=0.03$, in $\cal E$ and $\cal A$ polymerization schemes.
We should note that the characteristic strains in this figure  depict only the inspiral phase of binary mergers; there are two other phases after this particular phase. In this figure, a comparison between two polymerization schemes through the inspiral period shows that, the polymer correction in the $\cal E$ scheme will be amplified much more than the one in the $\cal A$ scheme,  which implies that the $\cal E$ scheme has more capacity to be observed in LISA. In these analyzes we have considered the largest possible values for the polymer parameters $\mu$ and $\nu$;  larger values than what already is reported here would generate waveforms with considerably different shapes and amplitudes, while the  corrections induced by the polymer effects are expected to be minuscule  compared to the classical waveform. For the localized sources of gravitational waves (e.g., black hole binaries), the scale $\ell$ can be set to the characteristic length of the given gravitational system.
In this regard, we have chosen $\ell = 10^{13}$m for the gravitational mode decompositions \eqref{eq:lambda-tot}, more comments about this point will be elaborated in the discussion section.

For the numerical analysis performed in this section, we need to know the mass and the distance of the binary system before hand. Usually GWs detectors are being used to find these parameters; if we want to compute analytically the expected strains of the classical and the polymer corrected GWs and compare them with detector's observation, other indirect methods can be used for finding these parameters, e.g., pulsar timing arrays \cite{Rosado:2015epa, Jenet:2003ew} or integral-field spectrograph \cite{Voggel:2022alp}.

\section{Discussion and Conclusions ~\label{sec:Discussion}}

GWs are the newest and one of the most important members of the multi-messengers currently used to explore the relativistic and quantum gravitational phenomena. These waves can potentially carry information about the early cosmology and the fine structure of spacetime, among others. 

In this manuscript we have considered a model of quantum GWs in polymer quantization scheme \citep{Garcia-Chung:2020zyq}, and have studied some of the consequences it may have on data taken by GW observatories, particularly LISA.  Using the geodesic deviation equation we have computed the modification to the classical dynamics and displacement of the detector hands of a GW observatory in such a model. We have studied both the fully nonperturbative/numerical, and also the analytical perturbative solutions to this modified dynamics. These in turn yield information about the modification to the GW strain registered by GW observatories. We find out that this model leads to the modification of the amplitude, frequency and the speed of propagation of the waves. These effects can be detectable by LISA if the polymer parameters have a minimum certain value. 

We also studied the coalescing waveform of two inspiral equal mass black holes. By analyzing the strain of this binary system in the Fourier (frequency) space, we have estimated the values of the polymer parameters needed so that LISA can observe the new physics corresponding to this quantum GW model. Numerical investigations in Sec.~\ref{sec:LISA} show that there is a possibility for observation of polymer corrections in certain conditions. Fig.~\ref{fig:CS} shows that closer binary systems with larger masses have more potential for observation of corrections induced by different schemes of polymer quantization. The analysis performed in Sec.~\ref{sec:LISA} demonstrates that the minimum detectable values for the polymer scales $\mu$ and $\nu$ in the given settings are  $10^{-50}\, \rm m^{1/2}$ and $10^{-58}\, \rm m^{-1/2}$ in natural units respectively. It should be emphasized that these values are found based on the presumption that polymer effects are sub-leading order effects and the overall shape of waveform should remain very close to the classical one. Thus, reported values are the largest possible values of $\mu$ and $\nu$ for a localized gravitational system with a given scale $\ell$ in which the polymer effects can be considered as corrections to the classical prediction in the given settings. According to the susceptibility range of LISA and LIGO/VERGO detectors for the frequency of GWs, values in range $10^{9} \rm m - 10^{13} \rm m$ for cutoff $\ell$ are acceptable when considering BHBs or super MBHBs. Choosing $\ell$ in this interval results in values of range $10^{-44} \rm m^{1/2} - 10^{-50} \rm m^{1/2}$ and $10^{-52} \rm m^{-1/2} - 10^{-58} \rm m^{-1/2}$ for polymer scales $\mu$ and $\nu$ respectively. While cosmological sources of gravitational waves can produce waves with wavelength up to the present cosmological horizon (or for the case of primordial gravitational waves, wavelength could be up to the size of last scattering surface), these wavelengths are more likely to be well beyond the reach of any direct detectors for the near future. The scenario in the present work is not cosmological, so the Hubble scale is not relevant for our context. Thus, in our setting it is physically reasonable to ignore or absorb such wavelengths in the homogeneous background. Here, we are dealing with localized gravitational systems, i.e., binaries. In the context of cosmological sources (like primordial GWs), we need to set $\ell$ to Hubble scale, this is what we will do in the upcoming paper in the context of inflation.

Colored curves in Fig.~\ref{fig:CS} depict only inspiral phase of binary merger, which means that if the polymer corrections in inspiral phase of the binary (dashed colored lines) are not in the sensitivity range of LISA, they may come in the detection range during the merger phase, due to the amplification of amplitudes, especially in polymer $\cal{E}$ scheme in which corrections increase more with frequency (compare slopes of dashed lines of Figs.~\ref{fig:CSE1} and \ref{fig:CSE2} with those in Figs.~\ref{fig:CSA1} and \ref{fig:CSA2}). 

In a future work, we will put a more strict bound on polymer parameters but performing a statistical analysis considering the data points from LIGO and comparing them with our theoretical results. One also can use the present results to compute the power spectrum of radiations. This we will also pursue in our next project.

\section*{ACKNOWLEDGMENTS}

Y.T. acknowledges the Research deputy of the University of Guilan for financial support.
This article is based upon work from the Action CA18108 -- Quantum gravity phenomenology
in the multi-messenger approach -- supported by the COST (European Cooperation in Science
and Technology). S. R. acknowledges the support of the Natural Science and Engineering Research Council of Canada, funding
reference numbers RGPIN-2021-03644 and DGECR-2021-00302.

\appendix

\section{Geodesic deviation}

Let us use the equations (\ref{def:action-Masses-2}) and (\ref{def:metric-fermi}) to derive the action for the geodesic deviation. According to these two equations, the action for the geodesic deviation takes the form
\begin{eqnarray}
S = - m \int dt \, \left\{ - g_{00}\left( \xi^j \right) - 2 g_{0i}\left( \xi^j \right) \dot{\xi}^i - g_{jk}\left( \xi^l \right) \dot{\xi}^j \dot{\xi}^k \right\}^{1/2}, \label{GDAction}
\end{eqnarray}
\noindent where the components of the metric tensor are given by (\ref{def:metric-fermi}) and at second order in $\xi^j$ are given by
\begin{eqnarray}
g_{00}\left( \xi^j \right) &=& 1 + R_{0i0j} \xi^i \, \xi^j, \label{MC1}\\
g_{0i}\left( \xi^j \right) &=& \frac{4}{3} R_{0jik} \xi^j \, \xi^k, \label{MC2} \\
g_{ij}\left( \xi^j \right) &=& - \delta_{ij} + \frac{1}{3} R_{ikjl} \xi^k \, \xi^l. \label{MC3} 
\end{eqnarray}

Replacing these coefficients in (\ref{GDAction}) yields
\begin{eqnarray}
S = - m \int dt \, \left\{ \left[ - 1 + R_{0i0j} \xi^i \, \xi^j + \left( \dot{\xi^i}\right)^2 \right] + {\cal O}(3, \xi) \right\}^{1/2}, \label{GDAction2}
\end{eqnarray}
\noindent where the Riemann coefficient term takes the form 
\begin{equation}
R_{0i0j} = - \frac{1}{2} \ddot{h}_{ij}(t,0).
\end{equation}
Assuming the analysis yields the action given in (\ref{GDAction-Ali}) let us continue from it and let us go over the extended phase space consideration. In other words, let us consider an action of the form
\begin{equation}
S = \int dt \left[ \frac{m}{2} (\dot{\xi}^i)^2 + \frac{m}{4} \ddot{h}_{ij}(t)\xi^i \, \xi^j \right]. \label{GDAction3}
\end{equation}
This action can be written as
\begin{eqnarray}
S &=& \int dt \left\{ \frac{m}{2} \left[ (\dot{\xi}^1)^2 + (\dot{\xi}^2)^2  \right] + \frac{m}{4} \left[ \ddot{h}_{11}(t) (\xi^1)^2 + 2 \ddot{h}_{12}(t)\xi^1 \, \xi^2 - \ddot{h}_{11}(t) (\xi^2)^2   \right] \right\}. \nonumber \\ \label{GDAction4}
\end{eqnarray}

\section{Canonical transformations}

In terms of new variables given by Eq.~(\ref{variables-new}), the action (\ref{GDAction-Ali}) takes the form
\begin{equation}
S_{\xi} \simeq \int_{\gamma_{B}} dt \left[ \frac{m}{2} (\dot{\xi}^{i})^2 +  \frac{m \kappa}{2\ell^{3/2}} \sum_{\lambda, \mathbf{k}} \ddot{\mathcal{A}}_{\lambda,\mathbf{k}}(t)\, e^{\lambda}_{jk}\, \xi^j \xi^k \right]. \label{GDAction-Ali1}
\end{equation}
or
\begin{align}
S_{\xi} &\simeq \int_{\gamma_{B}} dt \left[ \frac{m}{2} (\dot{\xi}^{i})^2 +  \frac{m \kappa}{2\ell^{3/2}} \sum_{\mathbf{k}} \left(\ddot{\mathcal{A}}_{+,\mathbf{k}}\, e^{+}_{jk}\, \xi^j \xi^k
+ \ddot{\mathcal{A}}_{\times,\mathbf{k}}\, e^{\times}_{jk}\, \xi^j \xi^k\right)  \right] \nonumber \\
& = 
\int_{\gamma_{B}} dt \left[ \frac{m}{2} (\dot{\xi}^{1})^2 + \frac{m}{2} (\dot{\xi}^{2})^2 +  
\frac{m \kappa}{2\ell^{3/2}} \sum_{\mathbf{k}} \left(\ddot{\mathcal{A}}_{+,\mathbf{k}}\, (\xi^1)^2 - \ddot{\mathcal{A}}_{+,\mathbf{k}}\,(\xi^2)^2
+ 2\ddot{\mathcal{A}}_{\times,\mathbf{k}}\, \xi^1 \xi^2\right)  \right] \nonumber \\
& = 
\int_{\gamma_{B}} dt \left[P_{\xi^1}\dot{\xi}^1 + P_{\xi^2}\dot{\xi}^2 - \left(\frac{P_{\xi^1}^2}{2m} + \frac{P_{\xi^2}^2}{2m}\right)  - 
\frac{m \kappa}{2\ell^{3/2}}  \left(\ddot{\mathcal{A}}_{+}\, (\xi^1)^2 - \ddot{\mathcal{A}}_{+}\,(\xi^2)^2
+ 2\ddot{\mathcal{A}}_{\times}\, \xi^1 \xi^2\right)  \right],
\label{action-arm2}
\end{align}
where, 
\begin{equation}
\dot{\xi}^i = \frac{P_{\xi^i}}{m}\, ,
\end{equation}
and we have defined
\begin{equation}
\mathcal{A}_{\lambda} \equiv \sum_{\mathbf{k}} \mathcal{A}_{\lambda, \mathbf{k}}.
\end{equation} 
By employing the  time-dependent canonical transformation 
\begin{align}
\left(\begin{array}{c} \xi^1 \\ \xi^2 \end{array} \right) = \left(\begin{array}{cc} P_{11} & P_{12} \\ P_{21} & P_{22} \end{array} \right) \left(\begin{array}{c} \chi^1 \\ \chi^2 \end{array} \right), \quad 
\left(\begin{array}{c} p_{\xi^1} \\ p_{\xi^2} \end{array} \right) = \left(\begin{array}{cc} P_{11} & P_{21} \\ P_{12} & P_{22} \end{array} \right) \left(\begin{array}{c} P_{\chi^1} \\ P_{\chi^2} \end{array} \right),
\end{align}
with the canonical map $\mathsf{P}$:
\begin{align}
\mathsf{P} = \left(\begin{array}{cc} P_{11} & P_{12} \\ P_{21} & P_{22} \end{array} \right)  = \frac{1}{\sqrt{2 \gamma}\, \ddot{\mathcal{A}}_{\times}} \left(\begin{array}{cc} (\gamma + \ddot{\mathcal{A}}_{+}) \sqrt{\gamma - \ddot{\mathcal{A}}_{+}} & (\gamma - \ddot{\mathcal{A}}_{+}) \sqrt{\gamma + \ddot{\mathcal{A}}_{+}} \\ (\gamma - \ddot{\mathcal{A}}_{+}) \sqrt{\gamma + \ddot{\mathcal{A}}_{+}} & -(\gamma + \ddot{\mathcal{A}}_{+}) \sqrt{\gamma - \ddot{\mathcal{A}}_{+}} \end{array} \right), 
\end{align}
where, $\gamma(t)$ is given by
\begin{align}
\gamma(t) = \sqrt{\ddot{\mathcal{A}}_{+}^2 + \ddot{\mathcal{A}}_{\times}^2}\, ,
\end{align}
we can rewrite the second term in the bracket in the action (\ref{action-arm2}) as
\begin{align}
\ddot{\mathcal{A}}_{+}\, (\xi^1)^2 
+ 2\ddot{\mathcal{A}}_{\times}\, \xi^1 \xi^2 - \ddot{\mathcal{A}}_{+}\,(\xi^2)^2 
&= 
\begin{pmatrix}
\xi^1 & \xi^2
\end{pmatrix} 
\begin{pmatrix}
\ddot{\mathcal{A}}_{+} & \ddot{\mathcal{A}}_{\times} \vspace{1mm} \\
\ddot{\mathcal{A}}_{\times} & -\ddot{\mathcal{A}}_{+}
\end{pmatrix}
\begin{pmatrix}
\xi^1 & \xi^2
\end{pmatrix}^{\rm T} \nonumber \\
&=: \begin{pmatrix}
\chi^1 & \chi^2
\end{pmatrix} 
\begin{pmatrix}
\gamma & 0  \\
0 & -\gamma
\end{pmatrix}
\begin{pmatrix}
\chi^1 & \chi^2
\end{pmatrix}^{\rm T}.
\end{align}

Now, in terms of the new canonical conjugate variables, i.e., $(\chi^i, P_{\chi^i})$ (it can be checked that $\{\xi^i, p_{\xi^i}\}=1=\{\chi^i, P_{\chi^i}\}$), the action (\ref{action-arm2}) becomes
\begin{align}
S_{\chi}
&= 
\int_{\gamma_{B}} dt \left[ P_{\chi^1}\dot{\chi}^1 + P_{\chi^2}\dot{\chi}^2 - \left(\frac{(P_{\chi^1})^2}{2m} + \frac{(P_{\chi^2})^2}{2m}\right) -  
\frac{m \kappa}{2\ell^{3/2}}  \gamma(t)\Big((\chi^1)^2 - (\chi^2)^2\Big)  \right] + \text{B.T.} \nonumber\\
&=: \int_{\gamma_{B}} dt \left[\left(P_{\chi^1}\dot{\chi}^1 - H_1\right) + \left(P_{\chi^2}\dot{\chi}^2 - H_2\right)\right] + \text{B.T.},
\label{HamilGDAction3a}
\end{align}
where,
\begin{align}
H_1 &:= \frac{(P_{\chi^1})^2}{2m} + \frac{m \kappa}{2\ell^{3/2}}  \gamma(t)(\chi^1)^2, \\
H_2 & := \frac{(P_{\chi^2})^2}{2m} - \frac{m \kappa}{2\ell^{3/2}}  \gamma(t)(\chi^2)^2.
\end{align}
It turns out that, the action (\ref{HamilGDAction3a}) describes two decoupled time-dependent harmonic oscillators, with Hamiltonians $H_1$ and $H_2$.
To obtain time-independent harmonic oscillators, we have to move to the extended phase space formalism. Of course, this have to be done for each oscillator independently of the other.

Now, we should look for equation of motion for the arm length and solve it.

\begin{align}
\dot{\chi}^1 &= \frac{\partial H_1}{\partial P_{\chi^1}} = \frac{1}{m} P_{\chi^1} , \qquad \dot{P}_{\chi^1} = - \frac{\partial H_1}{\partial {\chi^1}} = - \frac{m \kappa}{\ell^{3/2}}  \gamma(t) \, \chi^1, \\
\dot{\chi}^2 &= \frac{\partial H_2}{\partial P_{\chi^2}} = \frac{1}{m} P_{\chi^2} , \qquad \dot{P}_{\chi^2} = - \frac{\partial H_2}{\partial {\chi^2}} = \frac{m \kappa}{\ell^{3/2}}  \gamma(t) \, \chi^2,
\end{align}

The second order equations are given by
\begin{align}
\ddot{\chi}^1 = - \frac{ \kappa}{\ell^{3/2}}  \gamma(t) \, \chi^1, \qquad  \ddot{\chi}^2 =  \frac{ \kappa}{\ell^{3/2}}  \gamma(t) \, \chi^2.
\end{align}

\bibliography{References}

\end{document}